# Modeling the kinetic behavior of the Li-RHC system for energy-hydrogen storage: (I) absorption

Neves, A. M.[1, 2], Puszkiel, J.[2,3], Capurso, G.[2], Bellosta von Colbe, J. M.[2], Milanese, C.[4], Dornheim, M.[2], Klassen, T.[1, 2], Jepsen, J.[1, 2]

**Affiliation**

[1]Helmut Schmidt University (HSU), University of the Federal Armed Forces, Holstenhofweg 85, 22043 Hamburg, Germany

[2]Helmholtz-Zentrum Hereon, Institute of Hydrogen Technology, Max-Plank-Str. 1, 21502 Geesthacht, Germany

[3]IREC Catalonia Institute for Energy Research, 08930, Sant Adrià de Besòs, Barcelona, Spain

[4]Pavia H2 Lab, C.S.G.I. & Department of Chemistry, Physical Chemistry Section, University of Pavia, 27100 Pavia, Italy

Corresponding author e-mail: andre.neves@hereon.de

## Abstract

The Lithium-Boron Reactive Hydride Composite System (Li-RHC) (2 LiH + $MgB_2$ / 2 $LiBH_4$ + $MgH_2$) is a high-temperature hydrogen storage material suitable for energy storage applications. Herein, a comprehensive gas-solid kinetic model for hydrogenation is developed. Based on thermodynamic measurements under absorption conditions, the system's enthalpy $\Delta H$ and entropy $\Delta S$ are determined to amount to -34 ± 2 $kJ \cdot mol\ H_2^{-1}$ and -70 ± 3 $J \cdot K^{-1} \cdot mol\ H_2^{-1}$, respectively. Based on the thermodynamic behavior assessment, the kinetic measurements' conditions are set in the range between 325 °C and 412 °C, as well as between 15 bar and 50 bar. The kinetic analysis shows that the hydrogenation rate-limiting-step is related to a one-dimensional interface-controlled reaction with a driving-force-corrected apparent activation energy of 146 ± 3 $kJ \cdot mol\ H_2^{-1}$. Applying the kinetic model, the dependence of the reaction rate constant as a function of pressure and temperature is calculated, allowing the design of optimized hydrogen/energy storage vessels *via* finite element method (FEM) simulations.

## 1. Introduction

Mainly driven by the necessity of reducing the impact of fossil energy consumption on the environment, researchers have been looking for suitable alternatives for generation, storage and use of renewable energies. One of the most promising alternatives is the use of hydrogen as an energy carrier, which can significantly reduce the negative impact on the environment if the hydrogen is produced from renewable sources [1,2].

Current technology can produce hydrogen by several methods is able to produce [3] and convert it for utilization [4]. Not only hydrogen production but also its conversion and its storage are topics under intense investigation. Nowadays, a cost-effective, compact and safe system to store hydrogen represents a bottleneck for the broad implementation of hydrogen as a clean energy vector.

Today, physical storage methods are the most widely used technologies. While readily available, pressurized vessels and liquefied hydrogen have disadvantages involving inherent safety risks and the necessity of operating in extreme conditions (e.g., pressures up to 700 bar for gaseous or temperatures below the critical temperature -240.15 °C – or even below -252,15 °C considering ambient pressure – for liquid hydrogen), increasing the overall operation costs of such technologies [5]. A considerable amount of the energy stored in hydrogen is required for its liquefaction [6,7] or its compression to increase the stored hydrogen density [6]. Chemical storage methods, such as hydrides, are an alternative to the physical methods mentioned above and show technological potential since they can work under much milder conditions of pressure and temperature [5].

One of the main challenges to store hydrogen efficiently is related to the low volumetric energy density that can be achieved by the current hydrogen storage technologies; especially if mobile applications are envisioned. For hydrogen gas, the volumetric energy density is around 3 $kWh \cdot m^{-3}$ under STP conditions [8–10]. Even though this value can be improved by using the aforementioned physical storage methods, the use of hydrides offers the advantage of reaching a value of up to 150 kg $H_2 \cdot m^{-3}$ with the $Mg_2FeH_6$ complex hydride (equivalent to around 5000 $kWh \cdot m^{-3}$ [11]). By working under milder conditions, hydrides allow avoiding energy losses in compression in the case of gas-high pressure storage and boil-off in the case of cryogenic storage [12].

One of the highest hydrogen volumetric and gravimetric densities is found in the light-complex hydride $LiBH_4$, presenting a volumetric storage density of a theoretical value of 121 kg $H_2 \cdot m^{-3}$ [13,14]. $LiBH_4$ decomposes only at temperatures over 400 °C, releasing $H_2$, LiH and B as products. $LiBH_4$ has a theoretical hydrogen gravimetric capacity of 13.5 wt. %, considering its decomposition to LiH, B and $H_2$. This is because LiH is stable up to 900 °C [15]. Additionally, pristine $LiBH_4$ has rather poor reversibility, even under harsh conditions (over 400 ºC and 100 bar) [16].

In order to overcome these limitations, several approaches have been applied such as nanoconfinement, addition of transition metals, destabilization through different complexes and binary hydrides addition [17–21]. Among them, the so-called Reactive Hydride Composite (RHC) approach has been one of the most effective methods with potential for a practical application owing to its suitable hydrogen storage properties [9,17]. The use of boron compounds ($MgB_2$) instead of elemental boron to synthesize light borohydrides like $LiBH_4$, $NaBH_4$ and $Ca(BH_4)_2$ (among others) by gas-phase loading has been studied by Barkhordarian *et al.* [17]. By the combining 1 mol $MgH_2$ with 2 mol $LiBH_4$ a fully-reversible composite system could be obtained [22]. In this material (hereafter named Li-RHC), the theoretical absorption reaction under around 400 °C proceeds as follows [22,23]:

$$2\,LiH + MgB_2 + 4\,H_2 \rightarrow 2\,\mathrm{LiBH_4} + \mathrm{MgH_2} \qquad \textbf{(1)}$$

Under this assumption, the reaction proceeds in a one-step fashion, with the reactants' consumption considered to occur concomitantly. The overall absorption reaction is exothermal. It is however important to see that the decomposition of the stable $MgB_2$ (standard enthalpy of formation of $MgB_2$ is $\Delta H$ = -92.0 kJ·mol$^{-1}$) is endothermal [24]. Therefore, the overall enthalpy of reaction decreases, allowing the formation of $LiBH_4$, otherwise difficult to obtain. Thermodynamic calculations have shown that the theoretical reaction enthalpy of equation (1) amounts to -46 kJ·mol $H_2^{-1}$ [17]. Absorption and desorption behaviors are, however, markedly different for this material [25], what brings even more challenges for interpreting data and greatly limits the applicability of conclusions drawn from experimental investigations and theoretical calculations to their respective case-scenarios.

This reaction is generally accepted [22,23,25–32] as representative of the absorption process and proceeds as a one-step reaction [25,28]. However, it was first suggested by Vajo *et al*. [22] that between 400 and 450 °C a two-plateau region should exist. Cova *et al*. [23] have shown that after around 413 °C a two-plateau region can be seen. These plateaus are related to the equilibrium conditions of the $LiBH_4$ phase and the $Mg/MgH_2$ reaction. Although such expected transition has been observed in previous works, enthalpy and entropy values are seemingly in disagreement even for the single-plateau temperature range among different works [22,23,33]. The kinetic properties of the system are influenced fundamentally by temperature and pressure. However, they are also affected by many other factors, including, for example, the reacted fraction, use of additives (catalysts) [22,25–32,34], microstructure, particle size distribution [35], cycle number [9], and degree of compaction [36], among others.

Envisioning the application of Li-RHC in a hydrogen storage system, a complete kinetic and thermodynamic investigation is required. With this information, it is possible to develop kinetic models that describe the reaction rate of the material as a function of the reacted fraction, the operative pressure and the temperature. One of the main challenges in the design of hydrogen storage systems based on hydrides lays on the development of numerical models that describe the phenomena occurring upon hydrogenation and dehydrogenation. Trustable models allow to evaluate the practical feasibility of the system, reducing the time and costs associated with experimental evaluations. Furthermore, the numerical development can lead to the identification of the most relevant parameters, so that the design can be optimized, either by purely numerical methods or by its combination with novel computational approaches such as machine learning [37].

This work aims to develop a comprehensive kinetic model for the absorption reaction of Li-RHC (2 $MgH_2$ + $LiBH_4$) with 0.05 mol $TiCl_3$ as an additive. For this purpose, the assessment of the thermodynamic behavior is performed, as well as investigations of its kinetic properties. These results enabled the determination of the rate-limiting step of the hydrogenation process, the identification of the driving-force component and consequently, the calculation of apparent activation energy for the absorption reaction under a wide range of pressure and temperature conditions. Finally, the calculations allow the identification of a general equation that describes the kinetic behavior of the system for the chosen experimental conditions. To the best of our knowledge, this is the first time that a comprehensive kinetic model for the hydrogenation of Li-RHC is presented, which contributes with new insights for forthcoming investigations about the modeling and design of hydrogen-energy storage reservoirs.

## 2. Experimental

### *2.1. Material Preparation*

#### *2.1.1. High-Energy Ball-Milling (HEBM)*

For the preparation of the Li-RHC, LiH powder (Alfa Aesar, purity of ≥ 99.4 %) and $MgB_2$ powder (Alfa Aesar, purity of 99 %) were mixed in a 2:1 molar ratio. Then, 0.05 mol of $TiCl_3$ (Sigma Aldrich, purity ≥ 99.995 %) per mol of $MgB_2$ was added. The milling process was carried out in a planetary ball-milling device (Fritsch, Pulverisette 5, Germany) using a 76 mm diameter tempered steel milling vial (66 mm high) with 10 mm tempered steel balls under argon atmosphere, a BPR (ball mass to powder mass ratio) of 10:1, for a total milling time of 20 hours (4 hours milling, followed by 1 hour wait time, repeated 4 times), at 230 RPM, with 20% of volume filling (ball volume with respect to the vial internal volume). For each milling, an amount of 5 g of powder was inserted into the milling vial. These parameters were chosen based on analyses developed and thoroughly discussed in previous work [35]. The as-milled powder present the following characteristics: particle size ranging from 10 to 70 μm, surface area of about 15 $m^2/g$ and a rounded-platelet-like morphology [35].

All the handling was performed under argon atmosphere and the storage of the samples was done in a continuously purified argon-filled glove box (MBraun, Germany).

### *2.2. Kinetics and Thermodynamical Properties Assessment*

#### *2.2.1. Intrinsic Kinetic measurements, data handling and PCI curves*

The hydrogen absorption experiments to assess intrinsic kinetic* behaviors were performed using a Sieverts-type apparatus (HERA, Canada, Canadian Patent, Serial Number 2207149 [38]) equipped with a differential pressure sensor and calibrated volumes. The internal apparatus temperature was maintained at 40 ± 1 °C at all times. The sample heating was provided by an oven surrounding the whole sample holder. The thermocouple used to assess the temperature during the experiment was located on the sample holder's outer wall.

For these measurements, around 100-150 mg of material were used to assure isothermal and isobaric conditions, as well as homogeneous concentration changes in the mass of material. With this, the mass of material can be considered as a punctual mass, avoiding the influence of heat transfer and mass transport phenomena [39]. This condition is also taken in several works [9,25,27,34,48] applying gas-solid models to fit the experimental curves, in which the used mass ranges between 100 – 200 mg. The use of more mass can cause a deviation in the analysis of the intrinsic kinetic model, and mainly create a non-uniform temperature profile in the sample, so that one have neither isothermal nor homogeneous concentration changes in the used mass of material. Such condition leads to mismatch in the real intrinsic kinetic behavior, since the equilibrium pressure of the material in different parts would be different as the temperature changes, leading to a concentration profile in the material [39].

* "Intrinsic Kinetic" is here understood as the kinetics of hydrogen absorption reaction taken as a whole (with its many steps) that is, for every practical purpose, devoid of influence of heat management and mass flow limitations for the optimal proceeding of the reaction.

The experiments were performed starting at the 18$^{th}$ cycle to ensure that the material was already stabilized in terms of capacity and kinetic behavior (see also Figure S1 and Figure S2 of the Supplementary Material). The two sets of experiments were kinetic measurements at constant pressure (30 bar), with varying temperatures ranging from 312 °C up to 425 °C (10 measurements in regular intervals), and kinetic measurements at constant temperature (375 °C) with varying pressures, from 15 bar to 50 bar (with a step of 5 bar).

The acquired data was batch-processed with a specific Python (Python 3.7) script and further treated in OriginPro™ version 9.6.0.172 Software (Origin Lab Corporation). The fittings and statistical evaluation employed user-defined functions, which were added with the in-built tools of the program. The expression used for uncertainty propagation is shown in equation (S1) in the Supplementary Material.

The Pressure-Composition-Isotherms curves were acquired between 350 °C and 425 °C using a PCT Pro (SETARAM, Caluire, France) at the University of Pavia, Italy. A mass of approximately 150 mg was used. The amount of absorbed hydrogen was normalized as reacted fraction according to equation (2) for each of the curves individually.

$$\alpha = \frac{mH_{2,\ t}}{mH_{2,\ max}} \tag{2}$$

where $mH_{2,\,t}$ is the mass of hydrogen absorbed at a given time t after the experiment started, $mH_{2,\,max}$ is the maximum capacity reached at the time in which the experiment was terminated. Reacted fraction values are experimentally determined and always vary from 0 to 1.

To determine ∆H and ∆S *via* van't Hoff equation, the hydrogenation process of the studied material was considered as a single-step process. The start and end of the plateau region where taken, respectively, as 0.2 and 0.7. For these calculations, the mean value of the temperature $T_{mean}$ throughout the experiment was considered.

The ∆H and ∆S are explicitly considered with a negative sign, since the hydrogen absorption reaction is exothermic.

## *2.3. Kinetic Modeling*

### *2.3.1. Empirical Kinetic Model and General Definitions*

The approach used in the present work is the Separable Variable approach as described by equation (3) [40,41]. In this model, it is possible to obtain the variation of the reacted fraction α as a function of time t by determining three variables, namely, K(T), F(P) and G(α). The name of this method of deriving a kinetic model comes from its strategy, which implies: by keeping two (undetermined) variables constant, it is possible to determine the third one. By reorganizing the variables and keeping some of those constant, all of the variables can be determined.

$$\frac{d\alpha}{dt} = K(T) \cdot F(P) \cdot G(\alpha) \tag{3}$$

The G(α) term in equation (3) corresponds to the dependency of the reaction rate on intrinsic factors (defects, crystalline structure, etc.) and morphological changes of the particles (size and geometry) [42,43]. G(α) is represented by different expressions according to the gas-solid reaction model and as a function of the quantity reacted fraction α [40].

These different gas-solid kinetic models belong to different categories and each implies a rate-limiting step for the overall reaction progress. In Table 1, it is possible to see a description of the gas-solid models with their names, differential form G(α), integral form g(α), acronym taken in this work and corresponding references. There is a large number of gas-solid models, and thus only some of the best-fitting models considered in the present work as possible candidates are shown. For the complete table, please refer to Supplementary Material Table S1. For a more in-depth discussion of the application of these gas-solid kinetic models, please refer to the work of Puszkiel [40].

**Table 1 – Four of the best-fitting kinetic models**

| Model | Rate-Limiting Step | Differential form G(α) | Integrated Form $g(\alpha) = k * t$ | Acronym | Ref. |
|---|---|---|---|---|---|
| Johnson-Mehl-Avrami-Erofeyev-Kholmogorov, n = 1 | One-dimensional growth with interface-controlled reaction rate | $(1 - \alpha)$ | $-\ln(1-\alpha)$ | JMAEK, n = 1 | [9,25,40,41,44–48] |
| Johnson-Mehl-Avrami-Erofeyev-Kholmogorov, n = 1.5 | Three-dimensional growth of random nuclei with decreasing diffusion-controlled reaction rate | $\frac{3}{2}(1-\alpha)[(-\ln(1-\alpha)]^{\frac{1}{3}}$ | $[-\ln(1-\alpha)]^{\frac{2}{3}}$ | JMAEK, n = 1.5 | [34,40,47] |
| Contracting Area | Two-dimensional growth of contracting cylindrical volume | $2(1-\alpha)^{\frac{1}{2}}$ | $1-(1-\alpha)^{\frac{1}{2}}$ | CA | [40,41,47] |
| Contracting Volume | Three-dimensional growth of contracting sphere | $2(1-\alpha)^{\frac{1}{3}}$ | $1-(1-\alpha)^{\frac{1}{3}}$ | CV | [25,27,34,40,41,47] |

The kinetic constant can be expressed as a function of the temperature and pressure functionalities, as described in equation (4):

$$k(T,P) = K(T) \cdot F(P) \quad (4)$$

K(T) represents the temperature-dependent term. This functionality is defined by the Arrhenius form, as shown in equation (5).

$$K(T) = A \cdot \exp\left(\frac{-E_a}{R \cdot T}\right) \quad \textbf{(5)}$$

where A is the frequency factor (also called pre-exponential factor) and $E_a$ is the apparent activation energy (of the reaction occurring in the material as a whole).

The pressure-dependency component F(P) is related to the chemical reaction's driving force. This component includes the operative pressure P and the equilibrium pressure $P_{eq}$. The precise form of F(P) is not easily determined and it is usually tentatively chosen among possible candidates, which are functions of parameters mentioned above. For the sake of clarity, Table 3 in Section 3.2.2 shows the tentatively used expressions in the attempt of fitting the experimental data.

To calculate the F(P) values, the $P_{eq}$ expression was obtained from the entropy $\Delta$H and enthalpy $\Delta$S values by using the van't Hoff equation as described by equation (6) in Section 3.1. In all the calculations, the temperature and pressure were taken as the mean temperature $T_{mean}$ and mean pressure $P_{mean}$ throughout the course of the experiment.

A simplified description of the steps to perform such calculations is provided below.

1. A set of curves under the same $H_2$-pressure under different (but quasi-constant) T are collected. For each of these curves, K(T) and F(P) are assumed constant. By calculating the reacted fraction α from the experimentally acquired kinetic curves from 10 to 90 % of α it is possible to evaluate which reaction model G(α) best represents the kinetics of the studied system. The "reduced time method" proposed by Sharp *et al.* [49] and Jones *et al.* [50] was used for this determination. For additional information, please refer to the Supplementary Material;
2. Once a reaction model was assumed, a non-linear fit considering the integrated form g(α) is used to obtain individual sets of k(T, P) from the experimental data;
3. With these values of k(T, P), a first determination of the frequency factor A and the apparent activation energy $E_a$ is done by considering a reaction model G(α) and F(P) as constant for each curve. The values found here are not the final ones and new values are assumed in the following steps;
4. Assuming the calculated A, $E_a$, and the reaction model G(α), the different pressure-dependency term F(P) expressions are evaluated. After evaluation, if a suitable expression is found, new values for the frequency factor A and the apparent activation energy $E_a$ are assumed, thus, this is an iterative process;
5. With the determination of K(T), F(P) and G(α), it is now possible to perform the data validation by plotting the calculated and the experimental reacted fraction α against time t curves.

For the determination of a general reaction model expression, some remarks should be taken into consideration. First, the models presented here were originally developed for the modeling of nucleation and growth kinetics considering liquid-solid (solidification) and solid-solid interactions [44–46]. Later, these same models were adapted for gas-solid interactions for hydride kinetic modeling. In the present case, all the kinetic measurement temperatures are above 270 °C, which is the melting point for the $LiBH_4$ phase [28]. However, the fact that one of the phases is in the liquid state was not considered in this analysis. This is possible particularly because these models presuppose themselves the existence of a rate-limiting step for the reaction. The gas-solid model describes the changes of the materials upon hydrogenation and dehydrogenation based on the fitting of such models to experimental curves. An analysis of this complex system at the atomic level is out of the scope of this work. Moreover, the overall reaction, as shown in equation (1), considers that the reaction occurs as a single-step reaction. All these simplifications were taken and validated in previous works about the analysis of the rate-limiting step of this complex hydride system [9,25,48].

## 3. Results and Discussion

### *3.1. Thermodynamical Properties*

Figure 1 shows the hydrogenation pressure-composition isotherms measured in the range between 350 °C and 425 °C. The PCIs display a notable variation of the equilibrium pressure $P_{eq}$ with the amount of absorbed hydrogen (transformed fraction), which gives rise to a “sloped” plateau.

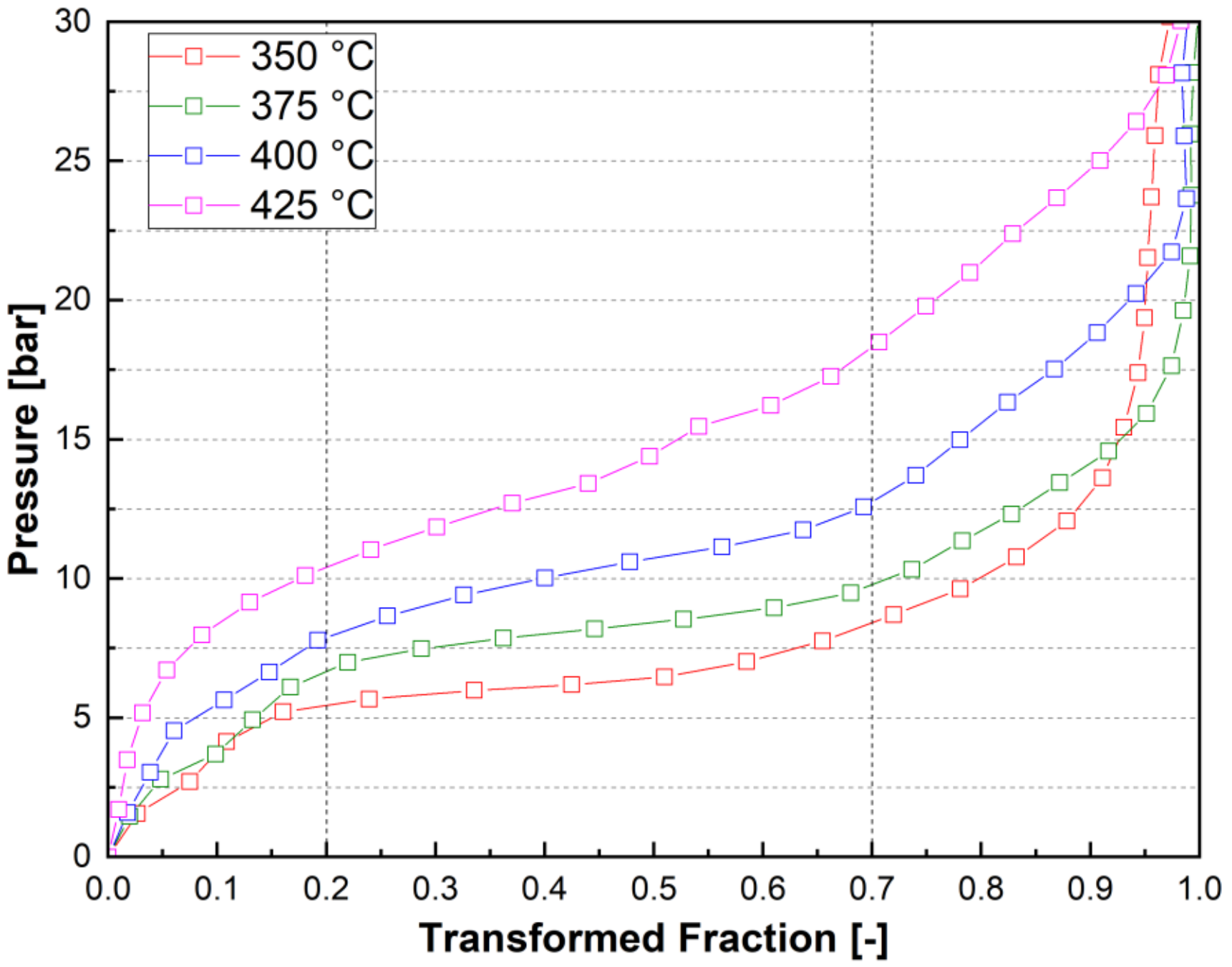

**Figure 1 - Pressure-composition-isotherm (PCI) curves for the studied Li-RHC at different temperatures under absorption conditions. The vertical lines indicate the points that were used for the thermodynamic parameter calculation.**

The thermodynamic parameters enthalpy ΔH and entropy ΔS can be calculated from the measured equilibrium pressure $P_{eq}$ by applying the van’t Hoff equation (6). At each temperature, the $P_{eq}$ was determined as a mean value between 0.2 and 0.7 of the transformed fraction α. Furthermore, a mean temperature $T_{mean}$ was taken considering the whole duration of the experiment. The $P_0$ is the thermodynamic reference pressure (considered 1 bar).

$$\ln\left(\frac{P_{eq}}{P_0}\right) = \left(\frac{\Delta H}{R \cdot T}\right) - \left(\frac{\Delta S}{R}\right) \quad \textbf{(6)}$$

By determining enthalpy ΔH and entropy ΔS values, it is possible to calculate an equilibrium pressure $P_{eq}$ as a function of the temperature T.

A comparative evaluation of the results for ΔH and ΔS and the pressure ranges reported in the literature is seen in Figure 2. The data points in Figure 2 were taken from the works of Vajo *et al*. [22], Cova *et al*. [23], Puszkiel [33] and this work. The values shown in the diagram are recalculated values as per reported in these works. Since only one of the works, namely, the one from Vajo *et al.*, reported the value for ΔS, for consistency, the recalculated values are instead being shown. Still, the differences between the recalculated values and the reported values were significantly different only for the single plateau for the work of Cova *et al.* [23]. A complete table with the used pressure values, the fitting parameters, the calculated and reported values is available in Table S2 in the Supplementary Material.

The straight lines represent the linear regression performed with the points shown in the diagram. The dashed line at around 412 °C is an approximation for the temperature range after which the reaction would not occur as a one-step reaction.

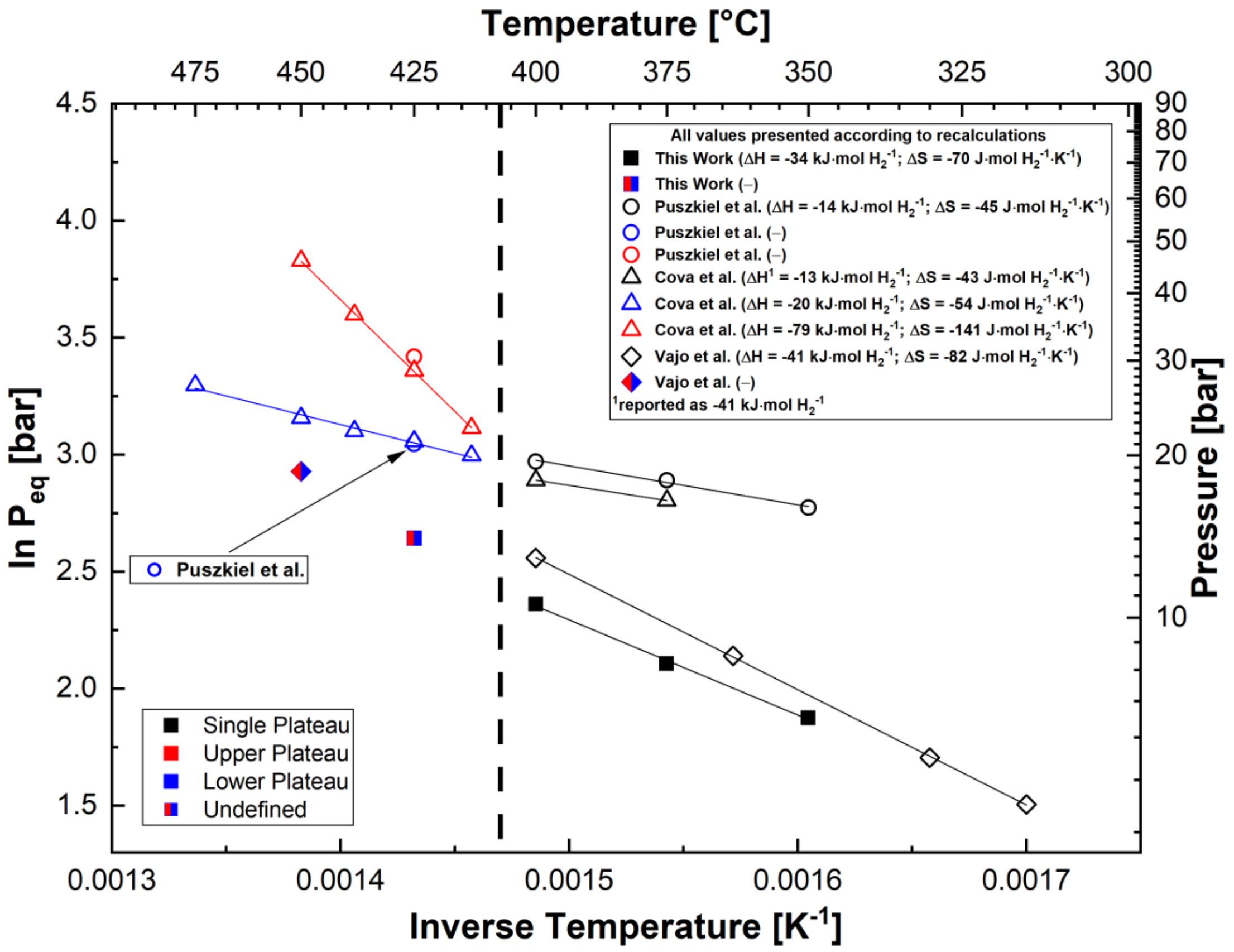

**Figure 2 – Comparative analysis for recalculated equilibrium plateau pressures for absorption reaction and the values of ΔH and ΔS from the cited works [22,23,33]*. Dashed line divides roughly the one-plateau from the two-plateau expected region.**

Vajo *et al.* [22] has reported that the ΔH and ΔS for this system amounts to -40.5 kJ·mol $H_2^{-1}$ and -81.3 J·$K^{-1}$·mol $H_2^{-1}$. Moreover, it has been proposed that the rise of a two-plateau reaction could possibly be observed for higher temperatures [22]. The data provided for the 450 °C PCI on that work could not resolve this issue. Later on, works from Puszkiel [33] and Cova *et al.* [23] have both been able to measure the presence of two plateaus for temperatures over 400 °C to 412 °C, respectively. In this range of temperatures, the system presents a region between the lower ($LiBH_4$) and upper plateau ($Mg/MgH_2$) in which it is theoretically possible to have coexistence of $LiBH_4$ and Mg in equilibrium conditions. This two-plateau behavior, however, is not clearly visible in the results presented in this work. This fact can be attributed to the differences in the equilibrium conditions resulting from differences in

* The reported value of the ΔH in their work of Cova *et al.* was of -41 ± 4 kJ·mol $H_2^{-1}$ for the single-plateau region. However, by taking the plateau points as described in their work, we were unable to reproduce the results. The points in the graph are being taken from the PCI readings, but as stated, both the recalculated and reported results are being shown here.

starting materials, handling, processing, among other experimental conditions. It should be expected that such a two-region plateau exists even in our material, but it may be possible that this transition occurs at higher temperatures.

As seen in Figure 3.a), for the hydrogenation process, the obtained values for $\Delta H$ and $\Delta S$ in this work are -34 ± 2 kJ·mol $H_2^{-1}$ and -70 ± 3 J·$K^{-1}$·mol $H_2^{-1}$ respectively. The obtained fitting goodness ($R^2$) of 0.994 shows a proper correlation. It is important to notice that the red-marked point at 425 ºC is not considered for the linear fitting. These values are in good agreement with the results published by Vajo *et al.* [22] for the temperature ranges below 400 °C. In Figure 1, for the curve at 425 °C, the “bump” seen around 0.55 and the steep increase in the pressure with increased reacted fraction suggests the presence of a second plateau. However, the difference observed here is not comparable to the change that has been reported in the work of Puszkiel [33] and Cova *et al*. [23]. Additionally, it should be noted that the point at 425 ºC (Figure 3.a)) presents a positive deviation over the fitted curve, in agreement with the work of Vajo *et al.*, where the positive deviation is seen at 450 °C [22]. This positive deviation can be ascribed to the presence of a two-plateau region. This issue is not still clear and would require further investigations that are beyond the scope of this work.

In the present work, for the sake of preciseness, a conservative assumption is made to guarantee the validity of the kinetic model in a range in which the mutual hydrogenation of $MgB_2$ and LiH to $LiBH_4$ and $MgH_2$ occurs. The temperature range of the kinetic data is limited to 412 °C. Above 412 ºC it may be possible that the system undergoes a different reaction pathway, and the kinetic analysis would not be representative. Therefore, the 425 °C kinetic results are, from now on, excluded from the kinetic modeling dataset.

With the calculated values of $\Delta H$ and $\Delta S$ it is possible to draw a pressure vs. temperature diagram (an exponential form of van’t Hoff equation) as seen in Figure 3.b). In this diagram, it is possible to see the equilibrium conditions for the Li-RHC system considering only the results found in the present study. Additionally, the points considered for the development of the kinetic model are shown as round markings.

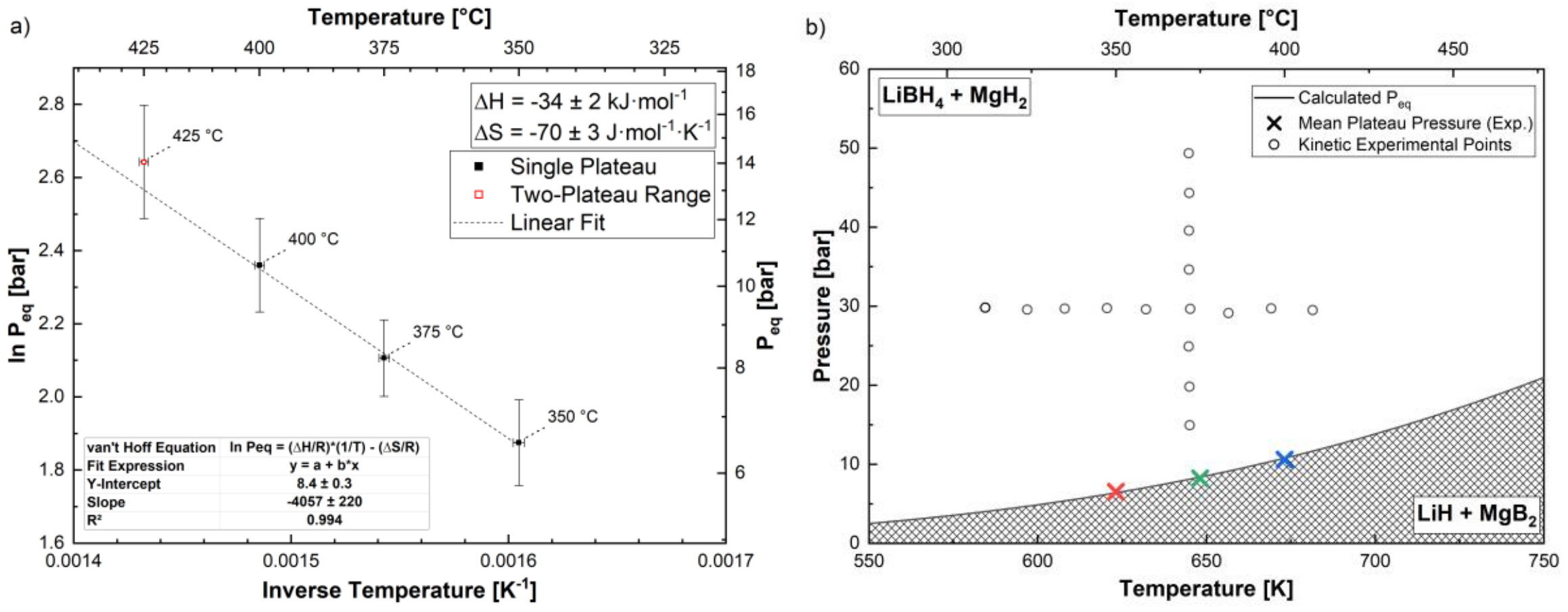

**Figure 3 – a) van’t Hoff plot for the PCI data, along with the fitted linear curve and its parameters. Note that the pressure at 425 °C was not used for the fitting. b) equilibrium absorption conditions in P vs T diagram, along with the measured kinetic experimental conditions. The x-markings represent the mean plateau pressure obtained experimentally in Figure 1.**

Additionally, the X-markings seen in Figure 3.b) are the experimentally measured mean plateau pressure values. It is possible to see that there are only small deviations between the values obtained from the van't Hoff equation and the experimental results used for the fit.

The general expression for calculating the equilibrium pressure $P_{eq}$ is given in equation 7) (van't Hoff Equation)

$$\ln(P_{eq}) = \left(\frac{-34 \cdot 10^3 \, J \, \cdot \, mol \, H_2{}^{-1}}{R \, \cdot \, T_{mean}}\right) - \left(\frac{-70 \, J \, \cdot \, mol \, H_2{}^{-1} \, \cdot \, K^{-1}}{R}\right) \quad \textbf{(7)}$$

where R is the ideal gas constant, and $T_{mean}$ is the mean temperature throughout the experiment.

### *3.2. Modeling of the kinetic behavior*

#### *3.2.1. Determination of G(α): Gas-Solid model*

Different reaction models have been developed to represent hydrogenation and dehydrogenation reactions. The models here considered for this purpose can be seen in Table S1 of the Supplementary Material. To deduce which of the reaction models best describes our system under the studied experimental conditions, the general approach is to start evaluating which of the integral equations g(α) (see Table 1) best correlates with the obtained experimental data. In order to do so, the so-called "Sharp and Jones" method (named after the original works of Sharp *et al*. [49] and Jones *et al*. [50] and also known as "Reduced Time Method") has been shown to be the most efficient tool, as it gives more parameters for evaluation that help determining how good is the agreement of the experimental data to each of the reaction models. A general description of the employed method has been given in the work of Puszkiel [40].

The evaluation of the best-fitting model is based on the values of three parameters; the coefficient of determination $R^2$ closest to 1, a Y-axis intercept closest to 0 and a slope closest to 1 [40,49,50]. All results are summarized in Table 2. Further information on how the method was implemented can be seen in the Supplementary Material, along with the figures for all the fittings done in this work (Figure S3).

All the fittings were performed for the measured temperatures within the range of 312 °C to 412 °C under 30 bar of initial $H_2$-pressure for each of the models. A graphical summary of the results in regards to the models in each temperature can be seen in Figure S4 of the Supplementary Material.

The obtained results indicate that there are 5 different models that rank sufficiently well in regard to $R^2$ values, namely, JMAEK with n = 1, JMAEK with n = 1.5, CV, 2-D diffusion limited and 3-D diffusion limited models. However, as slope and Y-intercept agreement are taken into account, both diffusion-limited models cannot be considered as suitable candidates. Regarding the three remaining models, JMAEK with n = 1.5 has insufficient proximity of the target values for slope and Y-intercept. Slightly closer to the targets is the CV model. Still, the JMAEK with n = 1 has, comparatively, the best match for the two parameters, especially considering Y-intercept.

**Table 2 – Fitting parameters for the three highest-ranking models for four chosen temperatures, under 30 bar of initial hydrogen pressure.**

| Model | Fit Parameter | 312 °C | 350 °C | 400 °C | 412 °C |
|---|---|---|---|---|---|
| JMAEK, n = 1 | $R^2$ | 0.996 | 0.999 | 0.999 | 0.999 |
| | Slope | 1.421 ± 0.002 | 1.183 ± 0.001 | 1.208 ± 0.001 | 0.968 ± 0.002 |
| | Y-Intercept | -0.369 ± 0.003 | -0.145 ± 0.002 | -0.073 ± 0.001 | 0.051 ± 0.004 |
| JMAEK, n = 1.5 | $R^2$ | 0.995 | 0.992 | 0.991 | 0.986 |
| | Slope | 0.841 ± 0.001 | 0.696 ± 0.002 | 0.717 ± 0.001 | 0.564 ± 0.005 |
| | Y-Intercept | 0.137 ± 0.002 | 0.277 ± 0.003 | 0.321 ± 0.002 | 0.401 ± 0.010 |
| Contracting Volume (CV) | $R^2$ | 0.995 | 0.993 | 0.991 | 0.987 |
| | Slope | 1.091 ± 0.002 | 0.903 ± 0.002 | 0.932 ± 0.001 | 0.735 ± 0.006 |
| | Y-Intercept | -0.097 ± 0.003 | 0.084 ± 0.004 | 0.139 ± 0.003 | 0.245 ± 0.012 |

Taking into account that the JMAEK with n = 1 model presents the highest $R^2$ values for all the temperatures (range from 0.996 to 0.999) and ranks rather well in the two other parameters, it is from now on assumed to be the most suitable model to describe the kinetics of the system.

For pristine Li-RHC, different authors reported differing models and even rate-limiting steps. In one of the first works related to the determination of rate-limiting steps and reaction models, Wan *et al.* claimed that the rate-limiting step should be the diffusion of species through the product layer [51]. Differently, Bösenberg *et al.* argued that an interface-limited model would be a better representation of the system [27]. Later on, Puszkiel *et al.,* in two different studies, argued in favor of an interface-controlled rate-limiting step [34,48]. Le *et al.* also reported an interface-controlled reaction [25]. Two models with interface-controlled kinetics were presented in these studies; the three-dimensional contracting-volume interface-controlled (3D CV) [27,34] and one-dimensional growth with interface-controlled reaction (JMAEK with n = 1) [25,48]. Although different models have been proposed, it should be noted that these models imply the same rate-limiting step. Furthermore, comparing these previous studies, some remarks should be taken into consideration. First, in some earlier works the JMAEK with n = 1 was not taken into account. Second, the method for determining and evaluating the fitting goodness of the kinetic models became more robust with time, since it became increasingly common practice to employ the Sharp and Jones method [40,48–50].

Concerning the Li-RHC with additives, different models were proposed for the system. Puszkiel *et al.* reported that, with the addition of 1 % mol of $TiO_2$, the model that best represents the system is the interface-controlled one-dimensional growth (JMAEK with n = 1) [48]. Le *et al.* reported that for the Li-RHC with the addition of 0.00625 mol of ($3TiCl_3{\cdot}AlCl_3$), the best-describing model changes in relation to the pristine material and becomes the 3D contracting-volume interface-controlled [25].

Particularly for $TiCl_3$, Bösenberg *et al.* claimed that by mixing this additive (the amount is not stated), for the second absorption at 350 °C and 50 bar, two possible models could be considered as suitable: the three-dimensional diffusion-controlled contracting-volume (3D CV diffusion-controlled) and the interface-controlled three-dimensional contracting-volume (3D CV interface-controlled) [27]. In their work, though, the JMAEK with n = 1 was not considered as a candidate. Additionally, the so-called reduced-time method based on the works of Sharp *et al*. [49] and Jones *et al*. [50] was not employed at the time.

In more recent studies for the 0.05 mol $TiCl_3$-added Li-RHC system it was found that the JMAEK with n = 1 best represents the experimental data [9]. This determination of g(α) was made based on different temperatures between 325 °C and 425 °C under 30 bar of initial $H_2$-pressure [9].

In the present work, by measuring the absorption kinetics in smaller steps (total of 10 measured curves between 312 °C and 425 °C) and by using the results after the 18$^{th}$ absorption cycle, it is reasonable to assume that both the influences of the calculation errors and of the change in hydrogen capacity with the first (de)hydrogenation cycles (altering the time to reach 90% of the reacted fraction) have been accounted for or reduced, respectively.

Considering that the JMAEK with n = 1 model describes a one-dimensional interface-controlled reaction, the results reported on [25,27,34,48] are in line with what is being proposed in this work.

The experimental results for reacted fraction α against time are shown in Figure 4.a) and Figure 4.b). The solid lines represent experimental kinetic data at different temperatures under 30 bar of initial $H_2$-pressure and for different pressures at 375 °C, respectively. The non-linear fit was performed to determine the k-values for the integrated JMAEK expression with n = 1 [44–46], as shown in equation (8):

$$\alpha = 1 - \exp(-k * t) \qquad \textbf{(8)}$$

The non-linear fitted curves are presented in Figure 4.a) and b) as dashed lines. Both graphs indicate clearly that higher pressures and higher temperatures – as usually expected – result in faster kinetics.

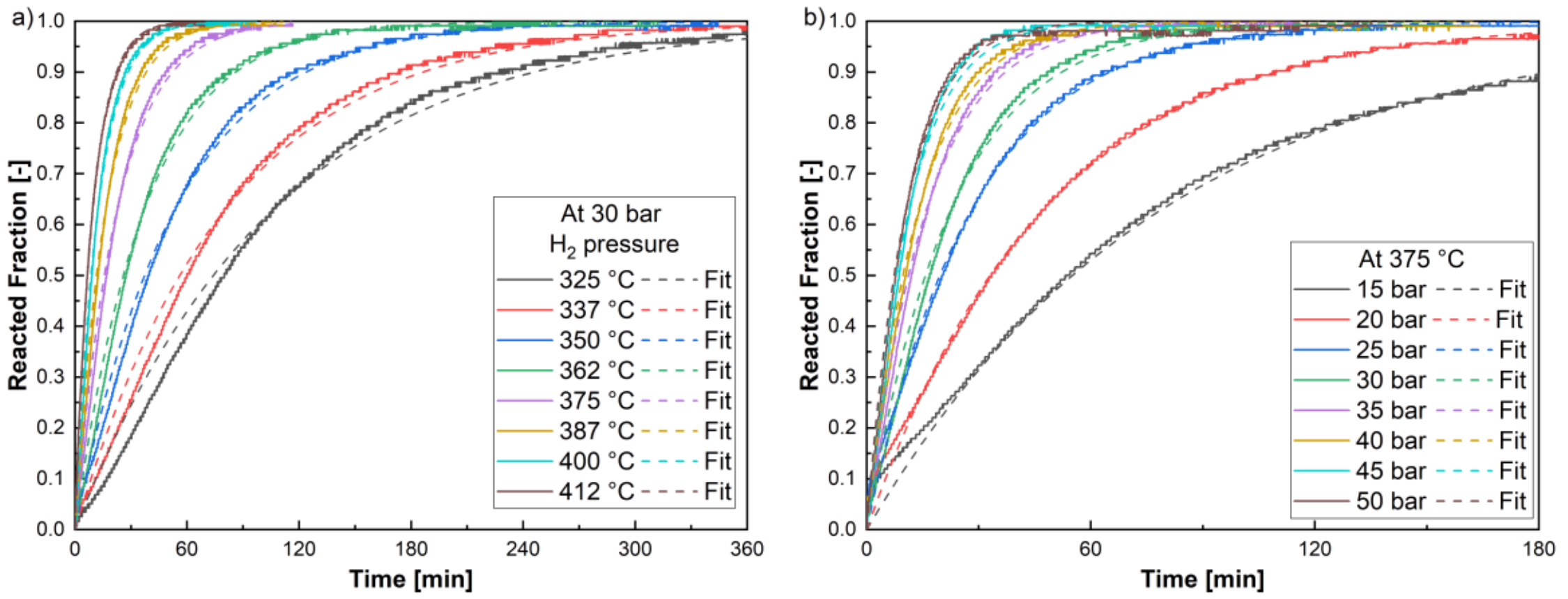

**Figure 4 – Experimental (solid lines) and fitted (dashed lines) curves for reacted fraction against time for a Li-RHC at a) different temperatures under 30 bar of initial $H_2$-pressure, b) at 375 °C under different pressures. Note that time scales are different for each of the plots.**

For the curves under a pressure of 30 bar, the non-linear fit performed represents well the experimental results obtained for most temperatures, with only minor observed deviations for the lowest temperatures. This deviation stems from the fact that the expression used for the fitting does not produce curves with inflections, which is observed clearly in the experimental curves for 325 °C and 337 °C in Figure 4.a). Such an aspect has been previously identified also in other works performed on this material [9]. The fitting parameters can be seen in Table S3 and S4 of the Supplementary Material.

A different kind of misfit between experimental data and fitted results is seen for the kinetic curves at 375 °C. For 15 and 20 bar pressures, the non-linear fits seemingly underestimate the hydrogen uptake during the first minutes of reaction. However, the increase in the reacted fraction for the experimental curves is likely related to an experimental artifact due to the time necessary to close the valve that initially sets the pressure difference between reference and sample holder volumes to zero. For a more in-depth description of the internal workings of the measurement apparatus, see [38]. It should be considered that the outline of the fitted curve presents a more realistic representation of the kinetics of the material under these experimental conditions than the experimental curve itself. Considering all the described effects, the deviations between the model and the experimental results are negligible. Therefore, it is possible to conclude that the proposed gas-solid model (equation 8) is in good agreement with the experimental results.

*3.2.2 Determination of temperature K(T) and pressure F(P) functionalities*

With the values from the k(T, P)-semi-empirical kinetic constants obtained from the non-linear fits of the experimental results from different temperatures under 30 bar initial $H_2$-pressure, the ln k against the inverse mean temperature are plotted in Figure 5.

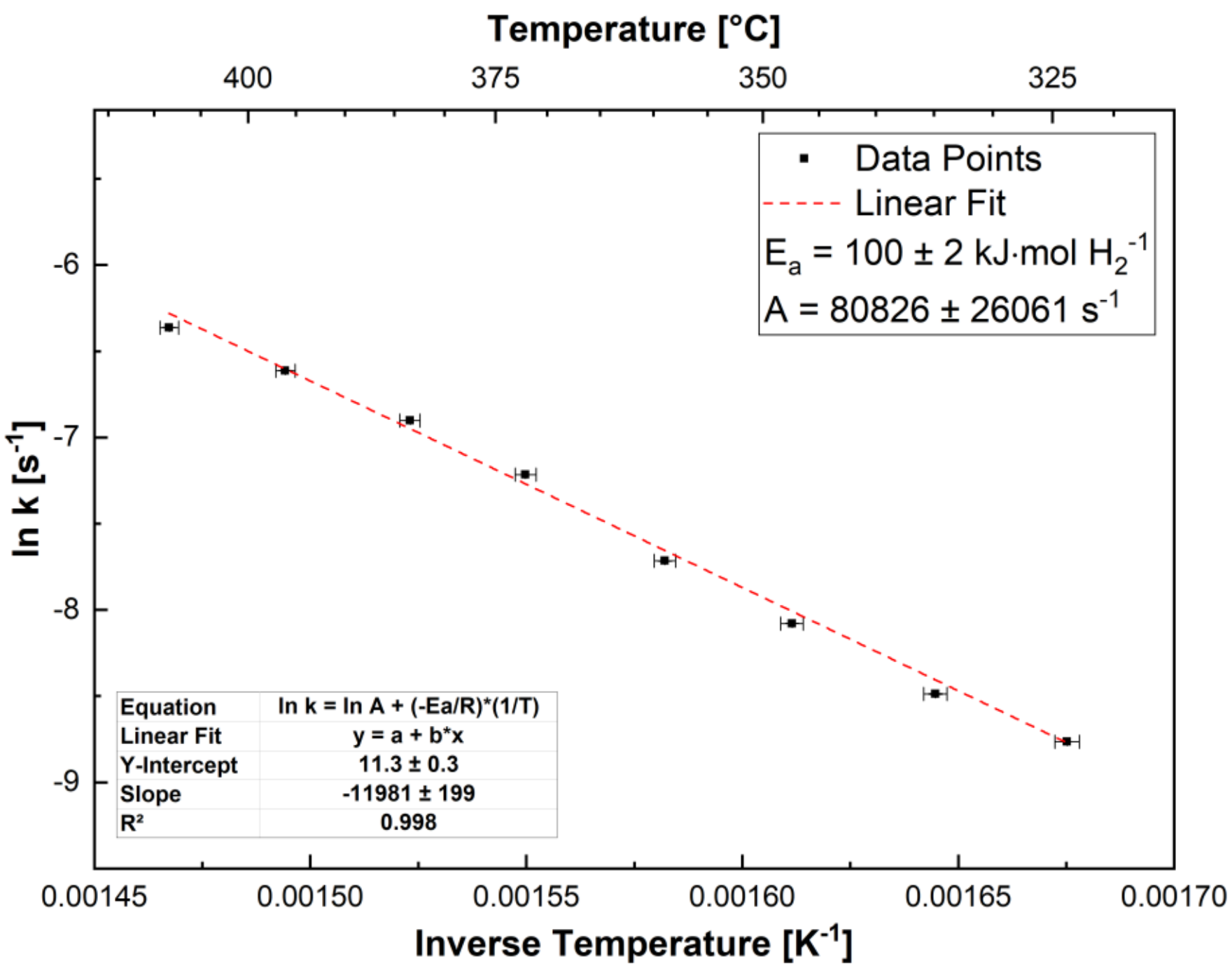

**Figure 5 – ln k against inverse mean temperature plot for different temperatures. The linear fit of the experimental data points is indicated as a dashed line. The absolute error bars for the Y-axis are not visible due to their small size.**

A linear fitting leads to the determination of the apparent activation energy $E_a$ and the frequency factor A. The linear fit presented a proper fitting goodness $R^2$ of 0.998. The values here found for the frequency factor A and the apparent activation energy $E_a$ were respectively 100 ± 2 kJ·mol $H_2^{-1}$ and $(8.1 \pm 2.6)\cdot10^4$ $s^{-1}$.

In order to take into consideration the influence of the experiment pressure P and the mean equilibrium pressure $P_{eq}$ at each temperature, the two found parameters, i.e., $E_a$ and A, are still to be corrected by the driving force component, F(P).

For the evaluation of the different models for F(P), equation (5) is divided into both sides by F(P) and combined with equation (4) resulting in

$$\frac{k(T,P)}{F(P)} = A \cdot \exp\left(\frac{-E_a}{R \cdot T}\right) \quad (9)$$

Now, by applying natural logarithms to both sides in (9) and rearranging, it becomes

$$\ln\left[\frac{k(T,P)}{F(P)}\right] = [\ln A] + \left[\frac{-E_a}{R}\right] \cdot \left[\frac{1}{T}\right] \quad (10)$$

With this linearized equation, it is possible to determine a frequency factor A and an apparent activation energy $E_a$ that take the pressure dependency into account by utilizing the values for k(T, P) (previously obtained from the fits shown in Figure 4), and each of the F(P) expressions. For the F(P) term, different functionalities are assumed, deducted either from investigations about the thermodynamic behavior or different tentatively proposed ratios and relations between P and $P_{eq}$. The expressions used in the present work are presented in Table 3.

**Table 3 – Summary of the F(P) expressions with conditions for utilization and/or conditions: mathematical expression, name used in this work and corresponding reference(s) .**

| Limited | Category/Conditions | Expression | Name | Reference |
|---|---|---|---|---|
| $F(P) > 0$ | $P = P_{eq} \Rightarrow F(P) \neq 0$ | $\frac{P}{P_{eq}}$ | F1 | [52] |
| | $P = P_{eq} \Rightarrow F(P) = 0$ | $P - P_{eq}$ | F2 | [53–56] |
| | | $\frac{P - P_{eq}}{P_{eq}}$ | F3 | [57] |
| | | $P^{0.5} - P_{eq}^{0.5}$ | F4 | [55,58] |
| | | $\left(\frac{P_{eq} - P}{P_{eq}}\right)^2$ | F5 | [59] |
| | | $ln\left(\frac{P}{P_{eq}}\right)$ | F6 | [56,60–63] |
| | $F(P)\ not\ a\ function\ of\ P_{eq}$ | $P$ | F7 | [64] |
| | $F(P)\ not\ a\ function\ of\ P_{eq}$ <br> $P_{eq} \approx 0$ | $P^{0.5}$ | F8 | [65] |
| $0 \leq F(P) < 1$ | $P = P_{eq} \Rightarrow F(P) = 0$ | $1 - \left(\frac{P_{eq}}{P}\right)^{0.5}$ | F9 | [59,66,67] |
| | $P = P_{eq} \Rightarrow F(P) = 0$ <br> $P_{eq} < P \leq 2P_{eq}$ | $\frac{\lvert P_{eq} - P \rvert}{P_{eq}}$ | -* | [68] |

To take the F(P) component into account, each expression is considered individually and a new plot, analogous to Figure 5, is built, along with each linear fit. This yields new values for $E_a$ and A for each of the F(P) functionalities, along with the $R^2$ values for each fitted curve.

Each new set of values for $E_a$ and A are individually considered (see also Supplementary Material Figure S5). By using the k-values obtained from the fittings shown in Figure 4 (see also Figure S6 and Figure S7 in the Supplementary Material), it is now possible to evaluate the different driving force expressions. Reorganizing (9), the expression

$$\left[\frac{k(T,P)}{A \cdot \exp(\frac{-E_a}{R \cdot T})}\right] = F(P) \quad \textbf{(11)}$$

can be obtained, for which the aforementioned values are applied.

Here, each F(P) expression (along with its results A and $E_a$) can be checked for its validity by plotting and interpreting it graphically. By analogy with a linear equation, the equation's left-hand side is taken as a dependent variable and F(P) as an independent variable. It follows that the best correlation would necessarily have a Y-intercept as near as possible to 0, a slope nearest to 1 and a determination coefficient $R^2$ closest to 1. For this particular fitting procedure, the linear coefficient is set to 0.

The results of this procedure for each F(P) expression can be seen in the Supplementary Material in Figure S6. The summarized results of these fitting parameters can be seen also in the Supplementary Material, in Figure S7. Comparatively, two F(P) expressions rank for $R^2$ much higher than the others, namely, F2 and F3, with $R^2$ values of around 0.990. These two best-ranking F(P) expressions are shown in Figure 6.a) and b).

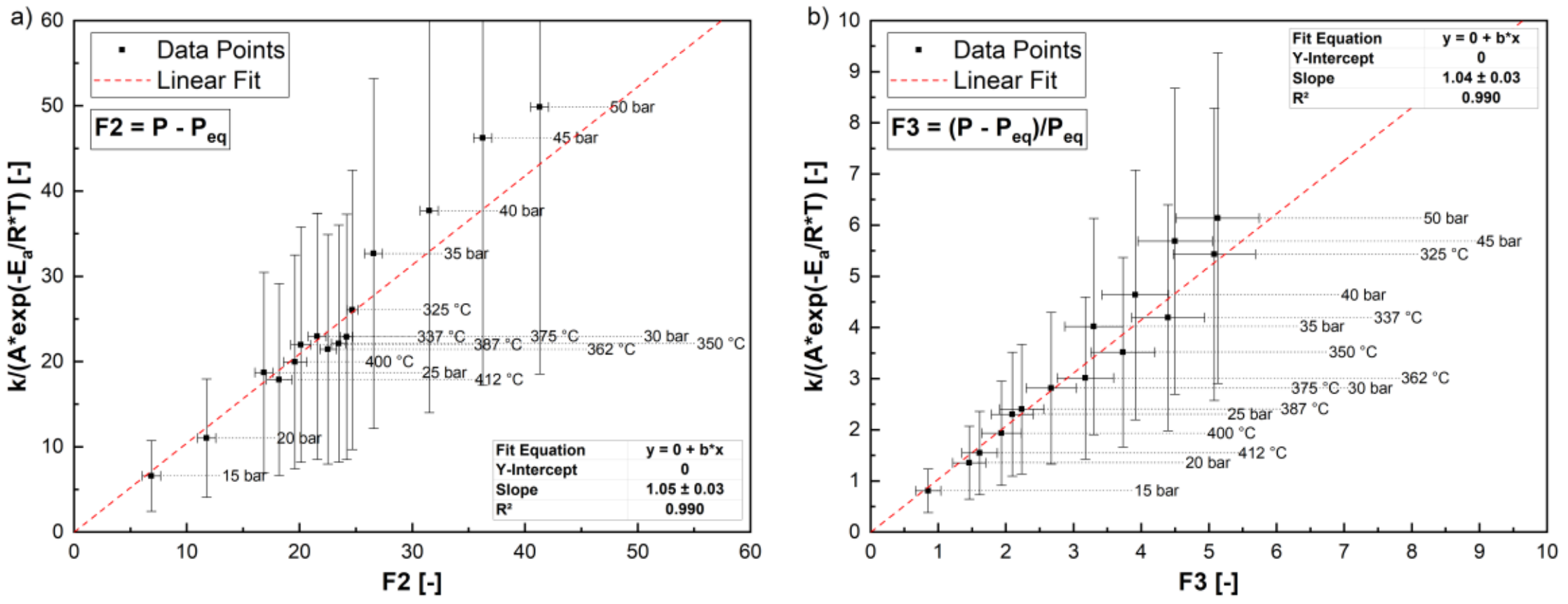

**Figure 6 - Fitting goodness verification for a) F2 and b) F3 driving force expressions.**

* This expression is numerically identical to F3 expression

The slope for F2 and F3 is respectively 1.05 ± 0.03 and 1.04 ± 0.03. These values are considered sufficiently close to 1, not only because of the method's uncertainty, but also for the comparative evaluation of the results.

However, just by comparing the parameters of the fitting, it is not possible to distinguish which one of the two expressions is the best choice for F(P).

By visual evaluation, it is possible to see that for F3 the deviations of the linear fit are more likely to be spread symmetrically around the fit curve. For F2, most of the deviations are for the high-pressure points and they all have a positive deviation.

As this deviation is more systematic in F2 than in F3, F3 is being favored, as it is less likely to present substantial deviations when calculated for their α values. Another aspect to take into account are the physical phenomena that were considered to propose said driving force expressions.

However, only limited information is currently available in the literature of why these expressions can fit the data of different pressures, as shown in our results. The driving force term of a chemical reaction is usually based on the thermodynamic activity. Considering the hydrogen gas as ideal, for a gas-solid reaction in a metal hydride, the thermodynamic activity is defined as the applied hydrogen partial pressure P divided by the standard pressure $P_0$ [69]. However, already for simple gas-solid hydride forming reactions, a wide variety of pressure dependence relations are used [58,59,68]. This fact is related to the different hydride forming materials, experimental conditions and rate-limiting steps associated to the F(P) [40].In the case of complex hydrides, it is well accepted that in general the relation between the applied pressure P and the equilibrium pressure $P_{eq}$ can describe the driving force term for the formation of a metal hydride. For instance, in the case of $NaAlH_4$, the functionality of the pressure was empirically determined, and the best function was determined as the first order Taylor series' approximation (centered around $P_{eq}$) of the change of the free energy for the hydrogenation process, i.e., $(P - P_{eq})/P_{eq}$ [57]. Therefore, in the herein investigated and rather complex $2\,LiBH_4 + MgH_2$ hydride system, the nature of the hydride forming materials, experimental conditions (set up) and the determined interface-rate limiting step leads to the $F(P)= (P - P_{eq})/P_{eq}$ as a best-fitting expression based on the above mentioned Taylor series' approximation. Due to its complexity, futher analysis regarding the physical meaning of the driving force term is beyond this work's scope.

The choice of this F(P) expression changes the previously-assumed apparent activation energy $E_a$ from 100 ± 2 to 146 ± 2 kJ·mol $H_2^{-1}$ and the frequency factor A from (8.1 ± 2.6) · $10^4$ to (1.8 ± 1.0) · $10^8$ $s^{-1}$.

### *3.2.3. General Expression and Data Validation*

Considering the calculated apparent activation energy $E_a$, the frequency factor A, the driving force expression F3, i.e., $F(P) = (P - P_{eq})/P_{eq}$, and the reaction model as JMAEK with n = 1, it is possible to propose a general expression for the studied system within the stated limits of temperature and pressure, i.e. from 325 °C to 412 ºC and from 15 bar to 50 bar. Starting from equation (3), substituting its components with K(T) as presented in equation (5), introducing the suitable expression for JMAEK with n = 1 (as seen in Table 1) and assuming F3 as the driving force expression, equation (12) is obtained, which is the differential form of the general kinetics expression.

$$\frac{d\alpha}{dt} = \left[\left(A * \exp\left(\frac{-E_a}{R\ *\ T}\right)\right) * \left(\frac{P - P_{eq}}{P_{eq}}\right) * (1\ -\ \alpha)\right] \quad (12)$$

Furthermore, by performing the integration of this expression, it is possible to obtain the integrated form of the equation, and by inputting the numeric values for the determined parameters, one can write the final formula of equation (13) as

$$\alpha\ =\ 1\ -\exp\left[-\left(1.8\cdot 10^8 * \exp\left(\frac{-146\cdot 10^3}{8.314\ *\ T}\right)\right) * \left(\frac{P - P_{eq}}{P_{eq}}\right) * t\right] \quad (13)$$

In this expression, the reacted fraction α varies between 0 and 1, the frequency factor A is given in ($s^{-1}$), the apparent activation energy $E_a$ is given in ($J\cdot mol\ H_2^{-1}$), the temperature T is given in (K), the pressure P is given in (bar) and the time t is given in (s).

By plotting equation (13) with the transformed fraction α against the time t, it is now possible to compare the calculated and the experimental kinetic curves. Figure 7.a) shows the validation plot for the experiments with different temperatures (in the 325 to 412 °C range) under 30 bar of initial $H_2$-pressure. In it, the solid lines represent the experimental data and the dashed lines, the calculated values. The insert on the bottom-right corner displays the same data but is limited to the higher temperature curves and the first 60 minutes of reaction.

For most of the temperature range, a very high degree of correlation between calculated and experimental results is achieved throughout the reaction. However, for the lowest temperatures investigated, e.g., 325 °C and 337 °C, the experimental curves present a perceptible inflection in the first hours, which is related to the significant complexity of the system reaction mechanism. This reaction modeling cannot capture this behavior, as the equations that are used imply a monotonic behavior and a curve that is always concave. Concerning these curves, it is visible that the kinetic reaction rate (curve slope) of the model is higher on the first minutes of reaction, slightly overestimating it when compared to the experimental data. After the inflection point, for most of the temperatures analyzed, the calculated reaction rates now are slightly underestimated. Before reaching the saturation point, the two curves again agree to a significant extent.

The coefficient of determination ($R^2$) for the calculated model and the experimental results was calculated according to equation (S2) of the Supplementary Material (from 0 to 0.99 reacted fraction). The values give a more objective frame of comparison between the curves' goodness-of-fit and possibly serve as a benchmark for other works for the future irrespective of the material used or reaction model proposed. Particularly for the absorption reactions at 375 °C all the $R^2$ values were above 0.996.

Although small visible differences between the model (equation 13) and the experimental results can be seen, it should still be considered that the applied model has been successful in describing the kinetics of the system, as these deviations can be considered minor in the frame of the course of the reaction. In Figure 7, it is possible to observe that the model shows a quite good agreement between the calculated and the experimental results with fitting goodness ranging between 0.97 and 0.99.

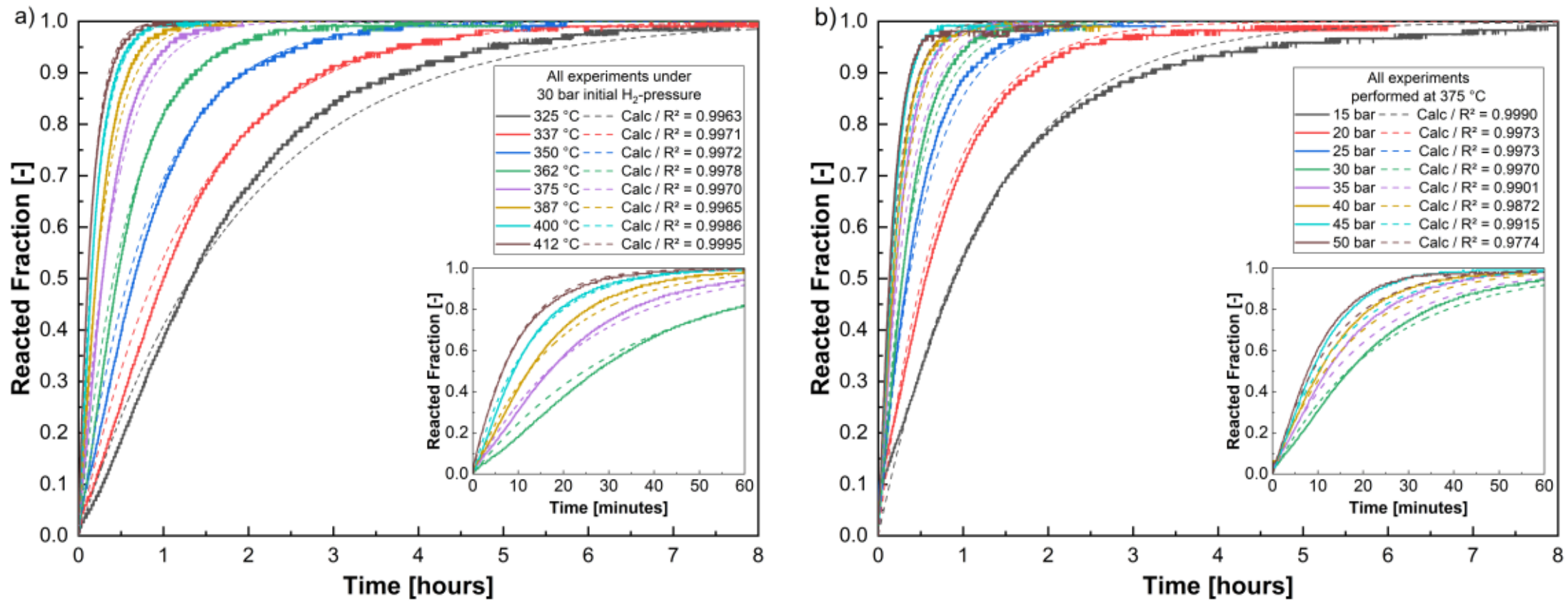

**Figure 7 - Comparison between calculated and experimental data for the variation of reacted fraction along the time for a) different temperatures at 30 bar initial $H_2$-pressure and b) different pressures at 375 °C along with the calculated coefficient of determination ($R^2$). The inset diagrams in the figures represent the same data under a shorter time scale.**

### *3.2.4 Thermodynamic stability and kinetic behavior: Isokinetic contour graphs*

An additional interpretation ofthe thermodynamic and kinetic results presented here is to draw isokinetic contour lines in a pressure-temperature diagram, as shown in Figure 8. The drawing of the isokinetic contour lines helps to understand how system conditions change influences the kinetic constant. For this purpose, a convenient approach is to represent the expression of the contour lines for different values of P as a function of T with chosen values of the kinetic constant k.

By rearranging equation (4), combining it with equation (5) as a function of temperature, making pressure the dependent variable, and implementing the F3 expression, it is possible to obtain the expression shown in equation (14). For plotting the isokinetic lines, k-values are arbitrarily chosen. Here, A is the frequency factor ($1.8 \cdot 10^8$ $s^{-1}$), the apparent activation energy $E_a$ (146 kJ·mol $H_2^{-1}$) and R is the ideal gas constant. The equilibrium pressure $P_{eq}$ is calculated by equation (7).

$$P = P_{eq} \cdot \left(1 + \left[\frac{k_{cte}}{A * \exp(-\frac{E_a}{R * T})}\right]\right) \quad \textbf{(14)}$$

This expression yields the contour lines seen in the inset graph of Figure 8 and in Figure S8 of the Supplementary Material. In both, the hatched region (below equilibrium curve) represents the equilibrium conditions that favor the stability of LiH + $MgB_2$, and the region above, in which the stable phases are $LiBH_4$ and $MgH_2$. This equilibrium line has been calculated using the van't Hoff equation (equation (7)) with the values of ΔH and ΔS obtained in the present work.

The curves drawn in the absorbed state region ($LiBH_4$ and $MgH_2$) are the isokinetic lines obtained from Equation (15). The values in Figure 8 were chosen to identify the experimental conditions in which at least two points are nearly intercepted by a single isokinetic line. In Figure S8, the values were chosen in order to show how the kinetic constant k(T, P) varies as a function of T and P.

The region in which the model is expected to reliably describe the kinetic behavior of the Li-RHC under absorption conditions is schematically shown in Figure 8 as a square limited by the temperature and pressure conditions for which this model is deemed valid. Thus, inside this region, the calculated isokinetic contour lines are solid, and outside the region, they are dashed lines, indicating that, in principle, the shape of these curves can be known with reasonable precision only inside this validity region.

The lower temperature border is around 325 °C, since, as shown in Table 2 (and in Figure S3 and Figure S4 of the Supplementary Material), the JMAEK with n = 1 does not fit well the kinetic curve at 312 °C. On the higher-temperature side (above 412 °C), the temperature range is limited by the change of the reaction pathway, resulting in different equilibrium conditions. In relation to the pressure limits, it is possible to assume that in the lower pressure range, if enough separation between the system conditions and the equilibrium condition ($P_{eq}$ curve) occurs, the model can still represent the kinetics of the system consistently. However, no experimental validation has been performed below 15 bar, not only because the times for the kinetics would be significantly large, but also because one of the goals of this work is to describe with enough detail the region (for conditions of P and T) that is of interest for potential applications of this material in energy storage systems. In the high-pressure region (above 50 bar), it is expected that the results obtained from the model can still represent the experimental results to some extent. However, as kinetics get faster in experimental conditions with a more significant driving force, it becomes increasingly harder to evaluate experimental results, as relatively small experimental uncertainties and experimental artifacts of measurement (valves opening/closing times and pressure transducer stabilization, *inter alia*) can lead to a significant change in the outcome and alter the interpretation of the results.

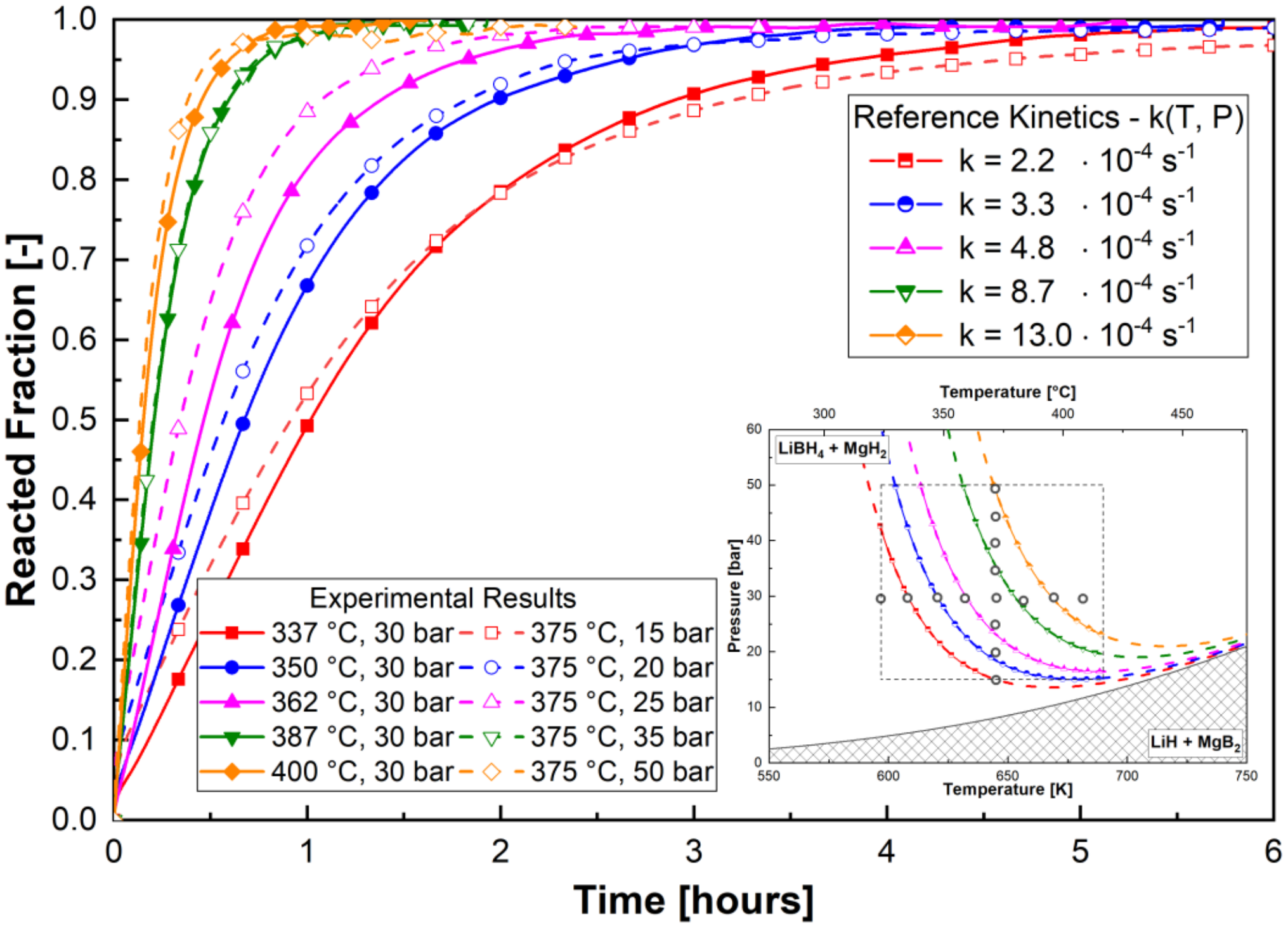

**Figure 8 – Kinetic model validation plot. In the inset diagram, the round circles represent each kinetic measurement at different pressures and temperatures. The delimited area marks the region of validity of**

**the kinetic model. The isokinetic curves that nearly intercept two experimental points are presented. In the main diagram, with the same colors, the experimental data for reacted fraction against time are presented.**

For each of the isokinetic curves drawn in the inset graph in Figure 8, the nearly intercepted experimental curves under these conditions are shown in the main graph. As these matching curves show very similar curve outlines, it has been shown that the found isokinetic line expression is able to reproduce the kinetic behavior of the material inside the aforementioned validity region quite well.

Still, it should be taken into consideration that the pressure and temperature ranges that delimit the validity of the model in this work are conservatively taken. That is, while the physical explanations provided to the kinetic model should hold true only in these ranges, extrapolation of the model to close-enough neighboring regions still yield excellent prediction capabilities (not shown here), showing the mathematical robustness of the model even for extended temperature and pressure conditions.

## 4. Conclusion and Outlook

In this work, both thermodynamic and kinetic experimental data are analyzed in detail for the $LiBH_4/MgH_2$ Reactive Hydride Composite (Li-RHC) with 0.05 $TiCl_3$ additive under absorption conditions. The obtained results in the range between 350 °C and 400 °C for $\Delta H$ and $\Delta S$ are -34 ± 2 kJ·mol $H_2^{-1}$and of -70 ± 3 J·$K^{-1}$·mol $H_2^{-1}$, respectively, in good agreement with previously reported values.

To the best of our knowledge, it is the first time that a comprehensive kinetic model under absorption conditions is developed for the Li-RHC by using the separable variable method. Applying this model, the effects of temperature, pressure and transformation of the hydride forming material are considered in the rate expression. The results indicate that the transformation of the forming hydride is limited by the movement of the not-hydrogenated/hydrogenated material interface, described by the one-dimensional interface-controlled model with a fixed number of nuclei and constant interface velocity, also known as JMAEK with n = 1 model. After taking the driving force component into account, the apparent activation energy $E_a$ and pre-exponential factor A are, respectively, (1.8 ± 1.0) · $10^8$ $s^{-1}$ and 146 ± 3 kJ·mol $H_2^{-1}$.

The developed model successfully describes the intrinsic kinetic behavior of the system, with only minor observed deviations. For the considered range of temperatures and pressures, the model shows fitting goodness ranging from 0.97 to 0.99. On-going research is being done to comprehensively describe the desorption kinetic behavior of the here studied material. Moreover, the developed models will be applied to model pilot-plant sized energy-hydrogen storage tanks through FEM simulations to better understand and optimize their designs.

The model presented here indicates that absorption times for the studied material are in the range of hours. While faster kinetics is a desirable trait for hydrogen storage materials, its influence on the performance of hydrogen storage systems depends heavily on the size and on the design choices for such systems [39]. For instance, intrinsic kinetics becomes increasingly unimportant as the size of the system increases. Still, the kinetic behavior is a core element of the functionally of these systems and needs to be properly described in order to obtain accurate descriptions of the combined effects that occur in these bigger-scale systems. However, faster kinetics have been demonstrated for Li-RHC in previous works. It has been previously demonstrated, that by producing ball-milled powders with some additives, it was possible to obtain nanostructures with in-situ-formed catalysts such as $Li_xTiO_2$ and AlTi compounds that enable the completion of the hydrogenation reaction with times of less than 30 minutes under 400 °C and 50 bar [25,48]. Our expectations are, that since in these cases the rate-limiting step has been also identified as interface-controlled, the modelling approach here presented can be similarly applied to describe these materials with improved kinetic behavior.

## 5. Acknowledgements

The authors would like to thankfully acknowledge the Karl-Vossloh-Stiftung for the financial support provided for this project (Project Number S047/10043/2017). Also, the authors acknowledge Kristin Przybilla for the experimental work related to the PCI assessment and for previous experiments, which have led to the current developments. Lastly, we thank Oliver Metz for all the technical support.

# Supplementary Material

## Modeling the kinetic behavior of the Li-RHC system for energy-hydrogen storage: (I) absorption

Neves, A. M.[1, 2], Puszkiel, J.[2,3], Capurso, G.[2], Bellosta von Colbe, J. M.[2], Milanese, C.[4], Dornheim, M.[2], Klassen, T.[1, 2], Jepsen, J.[1, 2]

Affiliation

[1]Helmut Schmidt University (HSU), University of the Federal Armed Forces, Holstenhofweg 85, 22043 Hamburg, Germany

[2]Helmholtz-Zentrum Hereon, Institute of Hydrogen Technology, Max-Plank-Str. 1, 21502 Geesthacht, Germany

[3]IREC Catalonia Institute for Energy Research, 08930, Sant Adrià de Besòs, Barcelona, Spain

[4]Pavia H2 Lab, C.S.G.I. & Department of Chemistry, Physical Chemistry Section, University of Pavia, 27100 Pavia, Italy

Corresponding author e-mail: andre.neves@hereon.de

**Table S1 – Gas-solid reaction models evaluated, with its description, differential form, integrated form, acronym used in this work and references for its description and/or use.**

| Model | Description | Differential form G(α) | Integrated Form $g(\alpha) = k * t$ | Acronym | References |
|---|---|---|---|---|---|
| Johnson-Mehl-Avrami-Erofeyev-Kholmogorov, n = 1 | One-dimensional growth of existing nuclei with constant interface-controlled reaction rate | $(1-\alpha)$ | $-\ln(1-\alpha)$ | JMAEK, n = 1 | [39, 40, 43–45] |
| Johnson-Mehl-Avrami-Erofeyev-Kholmogorov, n = 2 | Two-dimensional growth of existing nuclei with constant interface-controlled reaction rate | $2(1-\alpha)[-\ln(1-\alpha)]^{\frac{1}{2}}$ | $[-\ln(1-\alpha)]^{\frac{1}{2}}$ | JMAEK, n = 2 | [39, 40, 43–45] |
| Johnson-Mehl-Avrami-Erofeyev-Kholmogorov, n = 3 | Three-dimensional growth of existing nuclei with constant interface-controlled reaction rate | $3(1-\alpha)[-\ln(1-\alpha)]^{\frac{1}{3}}$ | $[-\ln(1-\alpha)]^{\frac{1}{3}}$ | JMAEK, n = 3 | [39, 40, 43–45] |
| Johnson-Mehl-Avrami-Erofeyev-Kholmogorov, n = 1.5 | Three-dimensional growth of random nuclei with decreasing diffusion-controlled reaction rate | $\frac{3}{2}(1-\alpha)[(-\ln(1-\alpha)]^{\frac{1}{3}}$ | $[-\ln(1-\alpha)]^{\frac{2}{3}}$ | JMAEK, n = 1.5 | [39, 40, 43–45] |
| Contracting Area | Two-dimensional growth of contracting cylindrical volume with constant interface velocity | $2(1-\alpha)^{\frac{1}{2}}$ | $1-(1-\alpha)^{\frac{1}{2}}$ | CA | [39, 40] |
| Contracting Volume | Three-dimensional growth of contracting cylindrical volume with constant interface velocity | $2(1-\alpha)^{\frac{1}{3}}$ | $1-(1-\alpha)^{\frac{1}{3}}$ | CV | [39, 40] |
| 1-D Diffusion | Surface-controlled reaction (Chemisorption) | $\frac{1}{2\alpha}$ | $\alpha^2$ | 1D | [39, 40] |
| 2-D Diffusion | Two-dimensional diffusion-controlled growth with decreasing interface rate | $-\frac{1}{\ln(1-\alpha)}$ | $((1-\alpha)\ln(1-\alpha))+\alpha$ | 2D | [39, 40] |
| 3-D Diffusion | Three-dimensional diffusion-controlled growth with decreasing interface rate | $\frac{3(1-\alpha)^{\frac{2}{3}}}{2\left(1-(1-\alpha)^{\frac{1}{3}}\right)}$ | $\left(1-(1-\alpha)^{\frac{1}{3}}\right)^2$ | 3D | [39, 40] |

3

## Uncertainty Propagation

The propagation of error was handled by applying

$$\sigma_f = \sqrt{\sum_{i=1}^{N}\left[\left(\left(\frac{\partial}{\partial x_n}[f]\right) \cdot \sigma_{x_n}\right)^2\right]} \tag{S1}$$

in which $f$ is a function of $N$ variables (e.g., $x_1, x_2, x_3$ ...), $\sigma_{x_n}$ is the best error estimate of $x_n$ and $\sigma_f$ is the best error estimate of the calculated value. It was assumed that the uncertainties of the variables are all independent of each other. Unless stated otherwise, the statistical uncertainties and error bars shown represent the standard error of the measured or calculated value.

## Coefficient of Determination ($R^2$) Calculation

The coefficient of determination ($R^2$) has been used in this work to assess how well the fitted or calculated curves could adjust the experimental results. For the fittings performed, the standard formulation of the software Origin Pro™ was used. For the calculation between the experimental data and the curves calculated from the model, equation (S2) was used

$$R^2 = \frac{\left(N \cdot \sum_{i=1}^{N}\left(y_i^{exp} \cdot y_i^{calc}\right) - \sum_{i=1}^{N}\left(y_i^{exp}\right) \cdot \sum_{i=1}^{N}\left(y_i^{calc}\right)\right)^2}{\left(\left(N \cdot \sum_{i=1}^{N}\left(y_i^{exp}\right)^2\right) - \left(\sum_{i=1}^{N}\left(y_i^{exp}\right)\right)^2\right) \cdot \left(N \cdot \sum_{i=1}^{N}\left(y_i^{calc}\right)^2 - \left(\sum_{i=1}^{N}\left(y_i^{calc}\right)\right)^2\right)} \tag{S2}$$

in which i is the index, N is the number of measurement points, $y^{exp}$ is the experimental value for reacted fraction and $y^{calc}$ is the calculated value for the reacted fraction.
To determine the $R^2$ of each of the curves in the different experimental conditions in which they were obtained, the reacted fraction between 0 and 0.99 was used.

**Table S2 – Data points used for construction of Figure 2 of the main manuscript file. Each line corresponds to a data point for a plateau pressure extracted from PCIs available for Li-RHC under absorption conditions. The obtained values for enthalpy and entropy are displayed, along with the values given in the original sources.**

| Source | Plateau | Temperature (K) | Inverse Temperature | Pressure (bar) | Slope | Y-Intercept | σ[Slope] | σ[Y-Intercept] | ΔH (mol $H_2$, calculated) | ΔS (mol $H_2$, calculated) | σ[ΔH] (mol $H_2$, calculated) | σ[ΔS] (mol $H_2$, calculated) | ΔH (mol $H_2$, reported) | ΔS (mol $H_2$, reported) | σ[ΔH] (mol $H_2$, reported) | σ[ΔS] (mol $H_2$, reported) |
|---|---|---|---|---|---|---|---|---|---|---|---|---|---|---|---|---|
| | | K | 1/K | bar | | | | | kJ/mol H2 | J/K · mol H2 | kJ/mol H2 | J/K · mol H2 | kJ/mol H2 | J/K · mol H2 | kJ/mol H2 | J/K · mol H2 |
| | | | | | ΔH/R | -ΔS/R | | | From Calculation | From Calculation | From Calculation | From Calculation | From Article | From Article | From Article | From Article |
| This Work | Single | 623,15 | 0,0016 | 6,52 | | | | | | | | | | | | |
| This Work | Single | 648,15 | 0,00154 | 8,22 | -4.070 | 8,3992 | 200,433 | 0,3097 | -34 | -70 | 2 | 3 | This Work | | | |
| This Work | Single | 673,15 | 0,00149 | 10,59 | | | | | | | | | | | | |
| | | | | | | | | | | | | | | | | |
| This Work | Undetermined | 698,15 | 0,00143 | 14,04 | - | - | - | - | - | | | | - | | | |
| | | | | | | | | | | | | | | | | |
| Puszkiel et al. [33] | Single | 623,15 | 0,0016 | 16,00 | | | | | | | | | | | | |
| Puszkiel et al. [33] | Single | 648,15 | 0,00154 | 18,00 | -1.663 | 5,4460 | 145,770 | 0,22524 | -14 | -45 | 1 | 2 | Not Reported | Not Reported | Not Reported | Not Reported |
| Puszkiel et al. [33] | Single | 673,15 | 0,00149 | 19,50 | | | | | | | | | | | | |
| | | | | | | | | | | | | | | | | |
| Puszkiel et al. [33] | Lower | 698,15 | 0,00143 | 21,00 | - | - | - | - | - | | | | - | | | |
| | | | | | | | | | | | | | | | | |
| Puszkiel et al. [33] | Upper | 698,15 | 0,00143 | 30,50 | - | - | - | - | - | | | | - | | | |
| | | | | | | | | | | | | | | | | |
| Cova et al. [23] | Lower | 686,15 | 0,00146 | 20,00 | -2.442 | 6,5477 | 154,414 | 0,21675 | -20 | -54 | 1 | 2 | -21 | Not Reported | 4 | Not Reported |
| Cova et al. [23] | Lower | 698,15 | 0,00143 | 21,25 | | | | | | | | | | | | |
| Cova et al. [23] | Lower | 711,15 | 0,00141 | 22,20 | | | | | | | | | | | | |
| Cova et al. [23] | Lower | 723,15 | 0,00138 | 23,50 | | | | | | | | | | | | |
| Cova et al. [23] | Lower | 748,15 | 0,00134 | 27,00 | | | | | | | | | | | | |
| | | | | | | | | | | | | | | | | |
| Cova et al. [23] | Upper | 686,15 | 0,00146 | 22,50 | -9.538 | 17,0152 | 106,212 | 0,15082 | -79 | -141 | 1 | 1 | -76 | Not Reported | 6 | Not Reported |
| Cova et al. [23] | Upper | 698,15 | 0,00143 | 28,75 | | | | | | | | | | | | |
| Cova et al. [23] | Upper | 711,15 | 0,00141 | 36,50 | | | | | | | | | | | | |
| Cova et al. [23] | Upper | 723,15 | 0,00138 | 46,00 | | | | | | | | | | | | |
| | | | | | | | | | | | | | | | | |
| Cova et al. [23] | Single | 648,15 | 0,00154 | 16,50 | -1.519 | 5,1462 | - | - | -13 | -43 | - | - | -41 | Not Reported | 4 | Not Reported |
| Cova et al. [23] | Single | 673,15 | 0,00149 | 18,00 | | | | | | | | | | | | |
| | | | | | | | | | | | | | | | | |
| Vajo et al. [22] | Single | 588,15 | 0,0017 | 4,50 | -4.927 | 9,8784 | 37,812 | 0,06073 | -41 | -82 | 0 | 1 | -40,5 | -81,3 | Not Reported | Not Reported |
| Vajo et al. [22] | Single | 603,15 | 0,00166 | 5,50 | | | | | | | | | | | | |
| Vajo et al. [22] | Single | 636,15 | 0,00157 | 8,50 | | | | | | | | | | | | |
| Vajo et al. [22] | Single | 673,15 | 0,00149 | 12,90 | | | | | | | | | | | | |
| | | | | | | | | | | | | | | | | |
| Vajo et al. [22] | Undetermined | 723,15 | 0,00138 | 18,70 | - | - | - | - | - | | | | - | | | |

## Kinetic Data Assessment

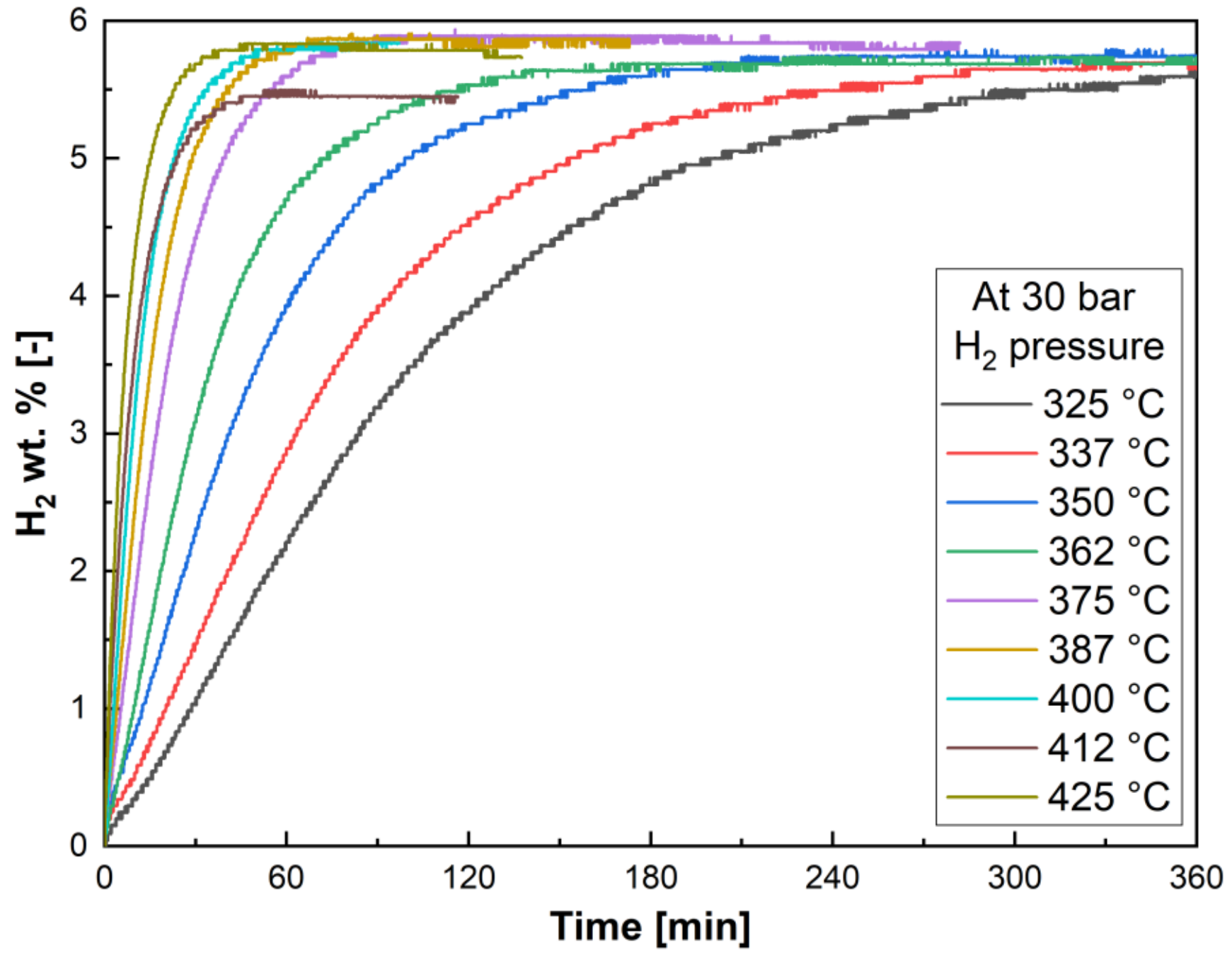

**Figure S1 – Kinetics of hydrogen uptake in weight percent for Li-RHC with added 0.05 $TiCl_3$ under 30 bar in a Sieverts apparatus under different temperatures (after at least 18 cycles).**

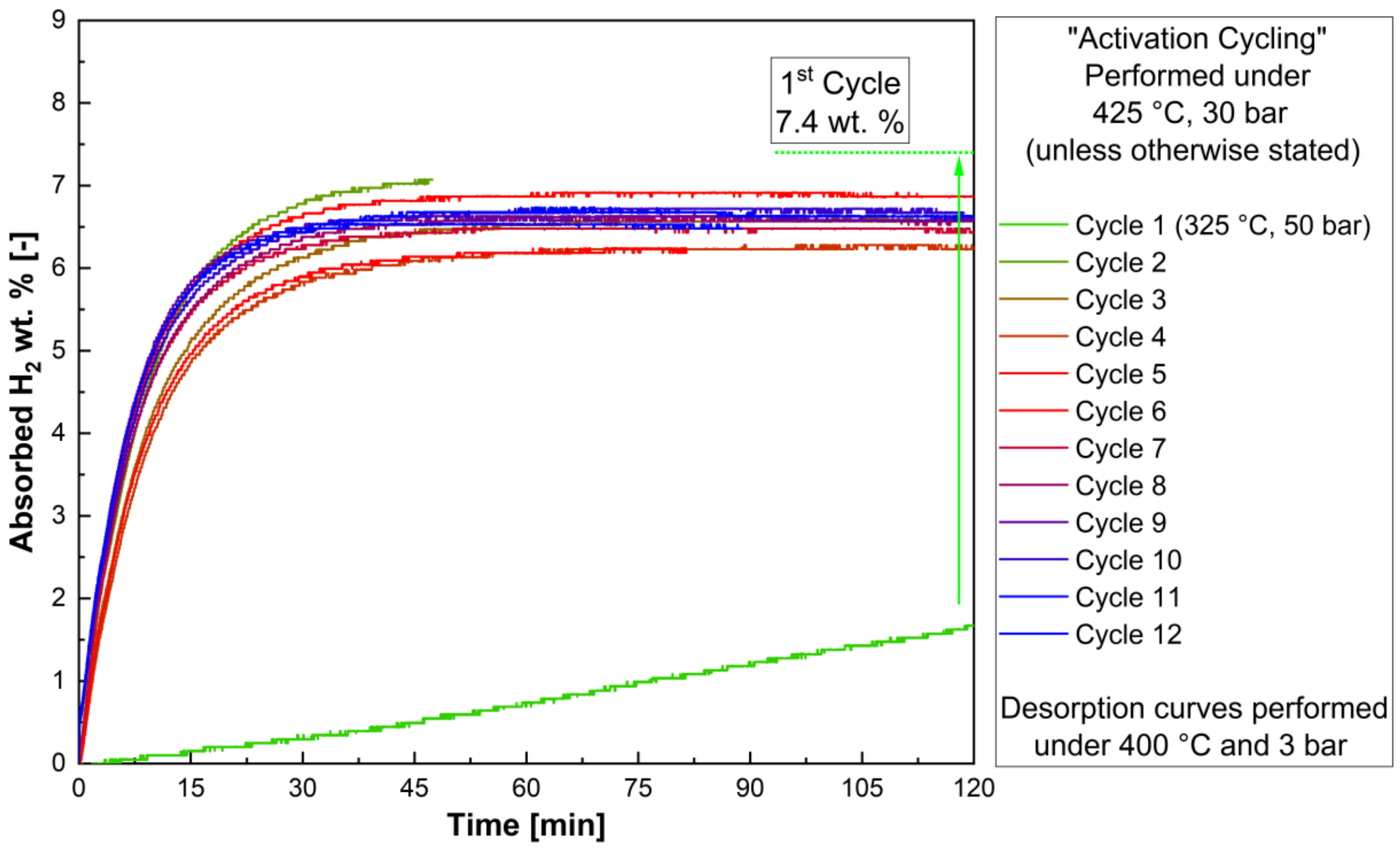

**Figure S2 – First 12 absorption kinetic curves for hydrogen uptake (envisioning capacity stabilization) in weight percent for Li-RHC with added 0.05 $TiCl_3$ under 30 bar in a Sieverts apparatus.**

**Additional Information on Methods: Sharp and Jones method implementation**

The so-called "Sharp and Jones method" is a mathematical data treatment strategy in order to ease the identification of the isothermal reaction kinetic model which would best describe experimental kinetic data. Such method has been popularized by the works of Sharp *et. al.* [48] and Jones *et. al.* [49] and is also known as "Reduced Time Method". Initially used for solid-state reaction kinetic model identification, it can also be applied also for gas-solid reaction kinetic models. The main advantage of such strategy is that it gives two additional parameters which can eventually help the evaluation of the results and improve the chances of correct selection of a suitable kinetic model to represent experimental data.

First, consider the integrated $g(\alpha) = k \cdot t$ expression (Table 1 of the main article file), in which α is the reacted fraction, k is a kinetic constant for a given temperature-pressure pair, t is the time and $t_{0.5}$ is the time a reaction takes to reach 0.5 of reacted fraction α. As the kinetic constant k does not change with the variation of t or α, it is possible to obtain equation (S3), which is independent of the value of k. As each reaction model has a determined g(α) expression, one can easily find the value of g(0.5) and $[t_{0.5}]_{theoretical}$.

$$\frac{g(\alpha)}{g(0.5)} = \frac{k \cdot t}{k \cdot [t_{0.5}]_{theoretical}} = \left[\frac{t}{t_{0.5}}\right]_{theoretical} \tag{S3}$$

Secondly, it is necessary to determine the value of $[t_{0.5}]_{experimental}$ (i.e., the value of the time t which was necessary to reach 0.5 of reacted fraction α). This can be done by plotting the experimental results of reacted fraction α against time t under the same pressure P in different temperatures T.

With $[t_{0.5}]_{experimental}$ and $[t_{0.5}]_{experimental}$ determined for each of the models, a plot of $[t/t_{0.5}]_{experimental}$ against $[t/t_{0.5}]_{theoretical}$ is made and the points are linearly fitted. The process is repeated for every temperature T and every model g(α). The plots are shown in Figure S3 for each of the studied temperature, under 30 bar of $H_2$-pressure, for each of the considered models.

A perfect correlation to the model is obtained when the fitted curve presents a coefficient of determination $R^2$ of 1, an Y-axis intercept of 0 and a slope of 1 [39, 48, 49].

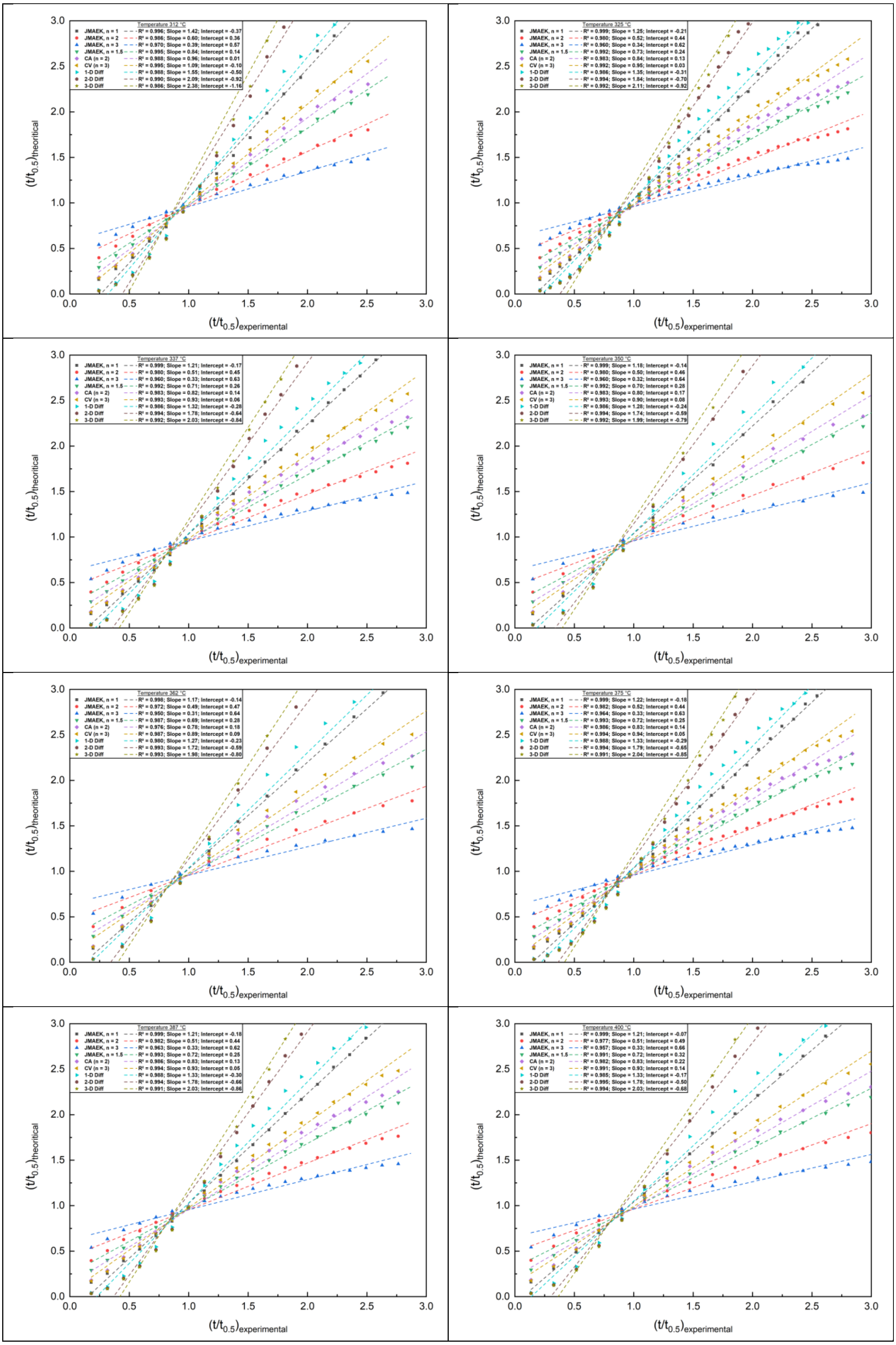
Temperature 312 °C
Temperature 325 °C
Temperature 337 °C
Temperature 350 °C
Temperature 362 °C
Temperature 375 °C
Temperature 387 °C
Temperature 400 °C
JMAEK, n = 1
JMAEK, n = 2
JMAEK, n = 3
JMAEK, n = 1.5
CA (n = 2)
CV (n = 3)
1-D Diff
2-D Diff
3-D Diff
$(t/t_{0.5})_{theoritical}$
$(t/t_{0.5})_{experimental}$

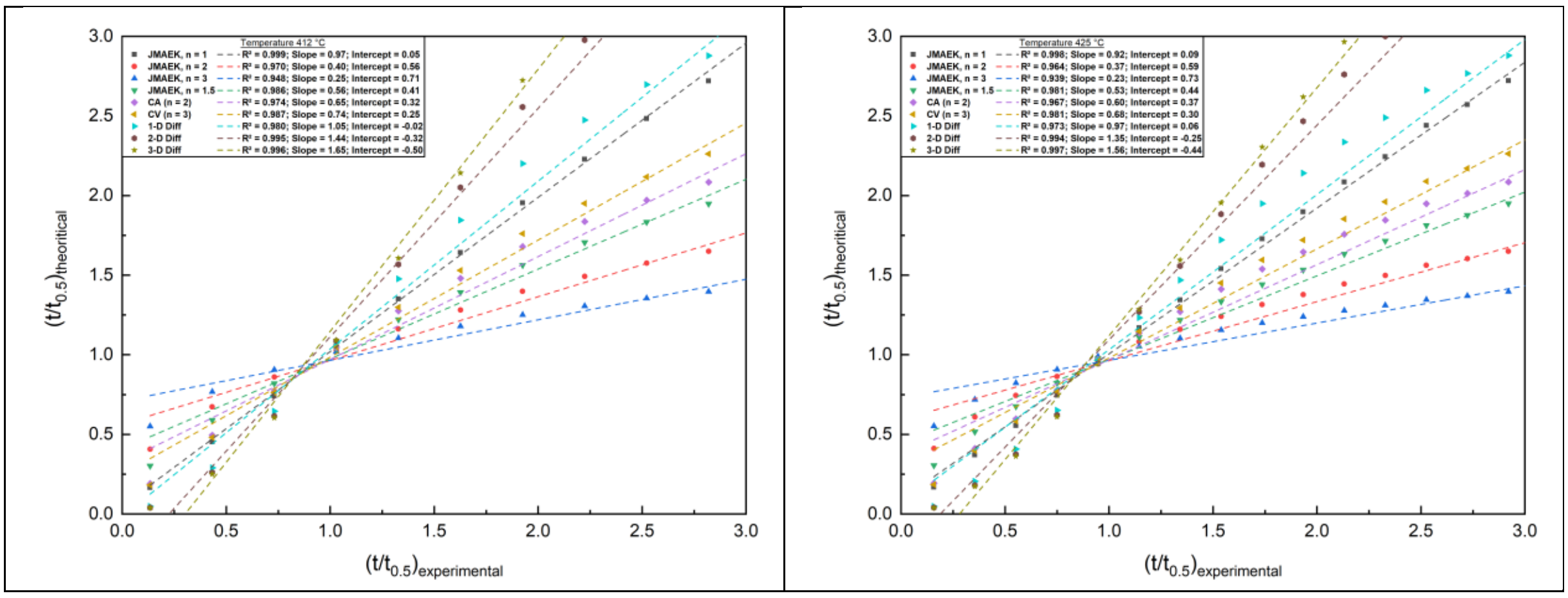

**Figure S3 – $[t/t_{0.5}]_{experimental}$ against $[t/t_{0.5}]_{theoretical}$ plots for the Li-RHC with 0.05 $TiCl_3$ for each of the temperatures under 30 bar of initial $H_2$-pressure.**

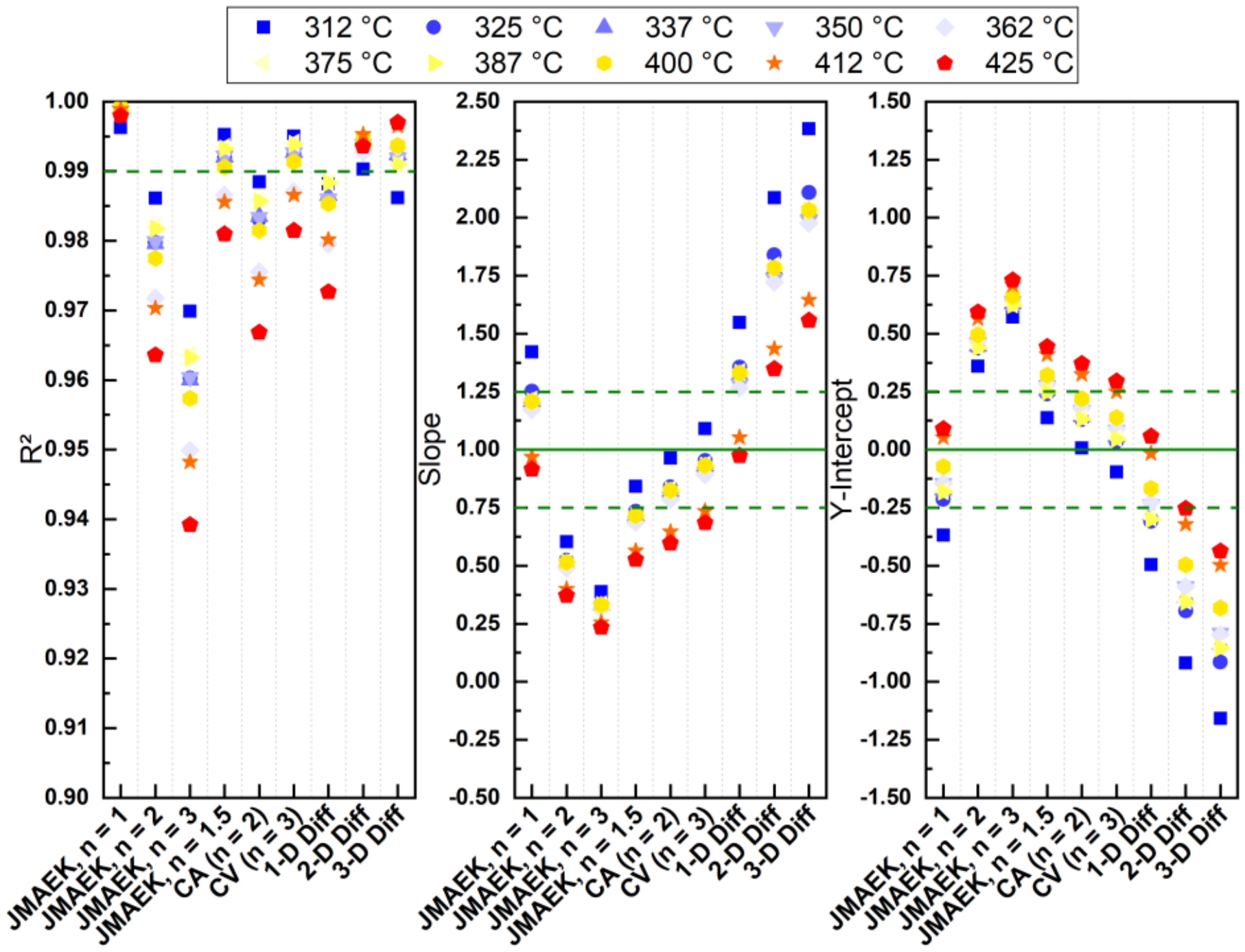

**Figure S4 –Results of the three fitting parameters (namely: coefficient of determination ($R^2$), slope and intercept) for all the studied temperatures, under 30 bar initial $H_2$-pressure for different models. The green lines represent the reference values for fit goodness and the dashed lines represent only a visual guide.**

## Kinetic Data Fit Parameters

**Table S3 – Non-linear fitting parameters of experimental data collected under 30 bar $H_2$-pressure under different temperatures (see Figure 4.a) of the main manuscript) indicating the used expression, the values for the kinetic constant k, its standard deviation σ(k) and the coefficient of determination $R^2$. Please, note that the values are being multiplied for a common factor (shown on the leftmost column) for clarity and to easy the reading.**

| Description | All experiments under 30 bar $H_2$-pressure | | | | | | | |
|---|---|---|---|---|---|---|---|---|
| Fit Equation | $\alpha = 1 - \left(exp(-k \cdot t)\right)$ | | | | | | | |
| Temperature [°C] | 325 | 337 | 350 | 362 | 375 | 387 | 400 | 412 |
| k [$s^{-1}$] ($k \cdot 10^3$) | 0.16 | 0.21 | 0.31 | 0.45 | 0.75 | 1.01 | 1.34 | 1.73 |
| σ(k) [$s^{-1}$] ($k \cdot 10^6$) | 0.04 | 0.25 | 0.37 | 0.39 | 0.52 | 0.75 | 0.90 | 2.24 |
| $R^2$ | 0.933 | 0.995 | 0.997 | 0.996 | 0.995 | 0.995 | 0.997 | 0.999 |

**Table S4 – Non-linear fitting parameters of experimental data collected under 30 bar $H_2$-pressure under different temperatures (see Figure 4.b) of the main manuscript) indicating the used expression, the values for the kinetic constant k, its standard deviation σ(k) and the coefficient of determination $R^2$. Please, note that the values are being multiplied for a common factor (shown on the leftmost column) for clarity and to easy the reading.**

| Description | All experiments at 375 °C | | | | | | | |
|---|---|---|---|---|---|---|---|---|
| Fit Equation | $\alpha = 1 - \left(exp(-k \cdot t)\right)$ | | | | | | | |
| Pressure [bar] | 15 | 20 | 25 | 30 | 35 | 40 | 45 | 50 |
| k [$s^{-1}$] ($k \cdot 10^3$) | 0.21 | 0.35 | 0.59 | 0.73 | 1.03 | 1.19 | 1.46 | 1.57 |
| σ(k) [$s^{-1}$] ($k \cdot 10^6$) | 0.24 | 0.38 | 0.72 | 0.45 | 2.33 | 2.61 | 5.40 | 5.22 |
| $R^2$ | 0.994 | 0.997 | 0.998 | 0.995 | 0.996 | 0.995 | 0.992 | 0.991 |

## $E_a$ and A determination after F(P) implementation

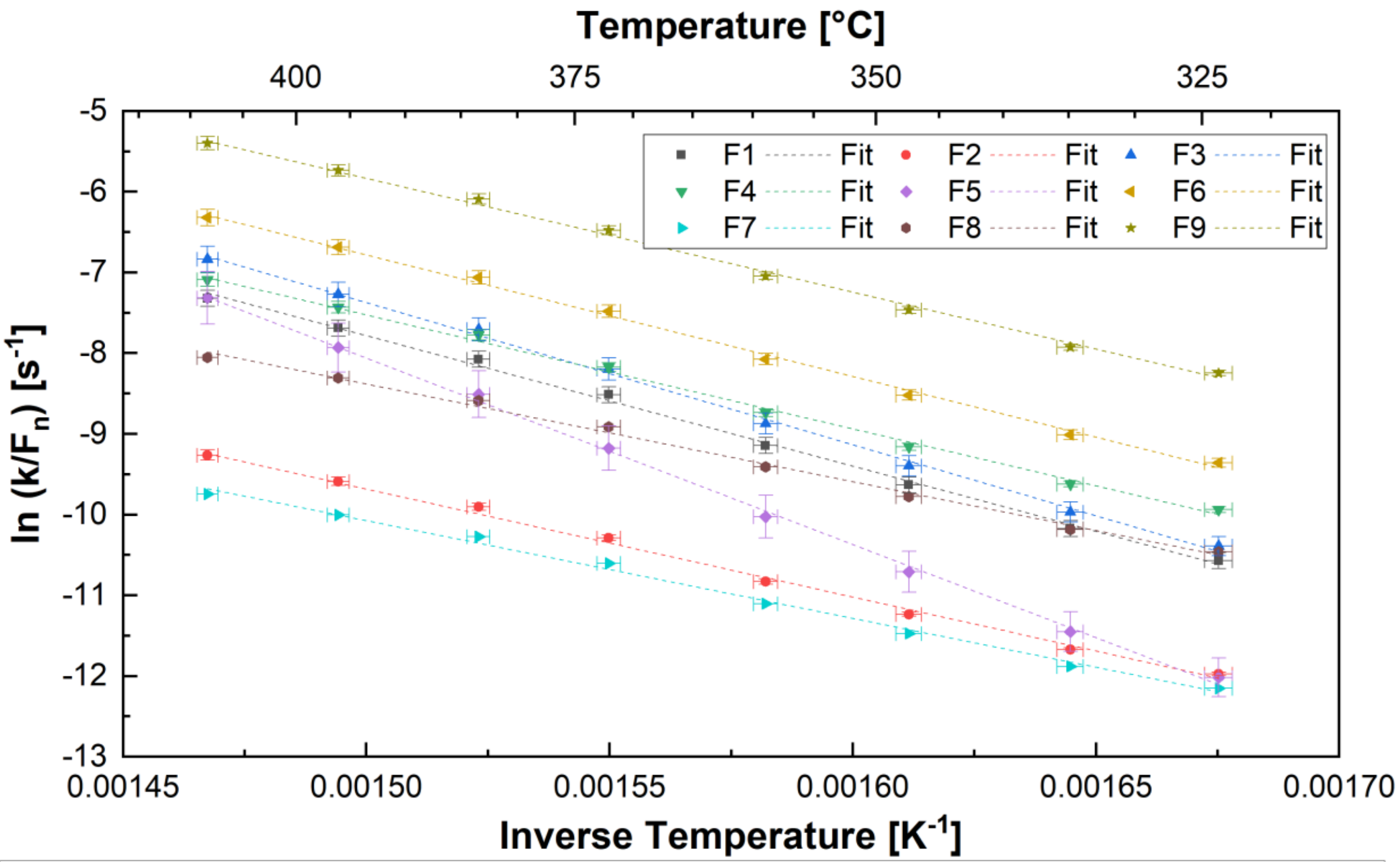

| ln (k/F) | | | | | | | | | |
|---|---|---|---|---|---|---|---|---|---|
| Named as | F1 | F2 | F3 | F4 | F5 | F6 | F7 | F8 | F9 |
| Expression | P/Peq | P - Peq | (P - Peq)/Peq | P^0.5 - Peq^0.5 | ((P - Peq)/Peq)² | ln(P/Peq) | P | P^0.5 | (1 - (Peq/P)^0.5) |
| Y-Intercept | 16.5 ± 0.5 | 10.4 ± 0.6 | 19.0 ± 0.5 | 13.6 ± 0.6 | 26.6 ± 0.6 | 15.8 ± 0.5 | 8.1 ± 0.5 | 9.8 ± 0.5 | 15.3 ± 0.6 |
| Slope | -16165 ± 330 | -13374 ± 389 | -17602 ± 332 | -14109 ± 364 | -23086 ± 385 | -15049 ± 343 | -12108 ± 330 | -12094 ± 320 | -14098 ± 353 |
| $R^2$ | 0.997 | 0.994 | 0.998 | 0.995 | 0.998 | 0.996 | 0.995 | 0.995 | 0.996 |

**Figure S5 – ln k/F(P) against the inverse of temperature plot for the Li-RHC with 0.05 $TiCl_3$ for the obtained data points and the linear fit for each of the driving force expressions used in this work. Below, a table with each of the expressions, the Y-intercept value, its standard deviation, the slope value, its standard deviation and the coefficient of determination $R^2$.**

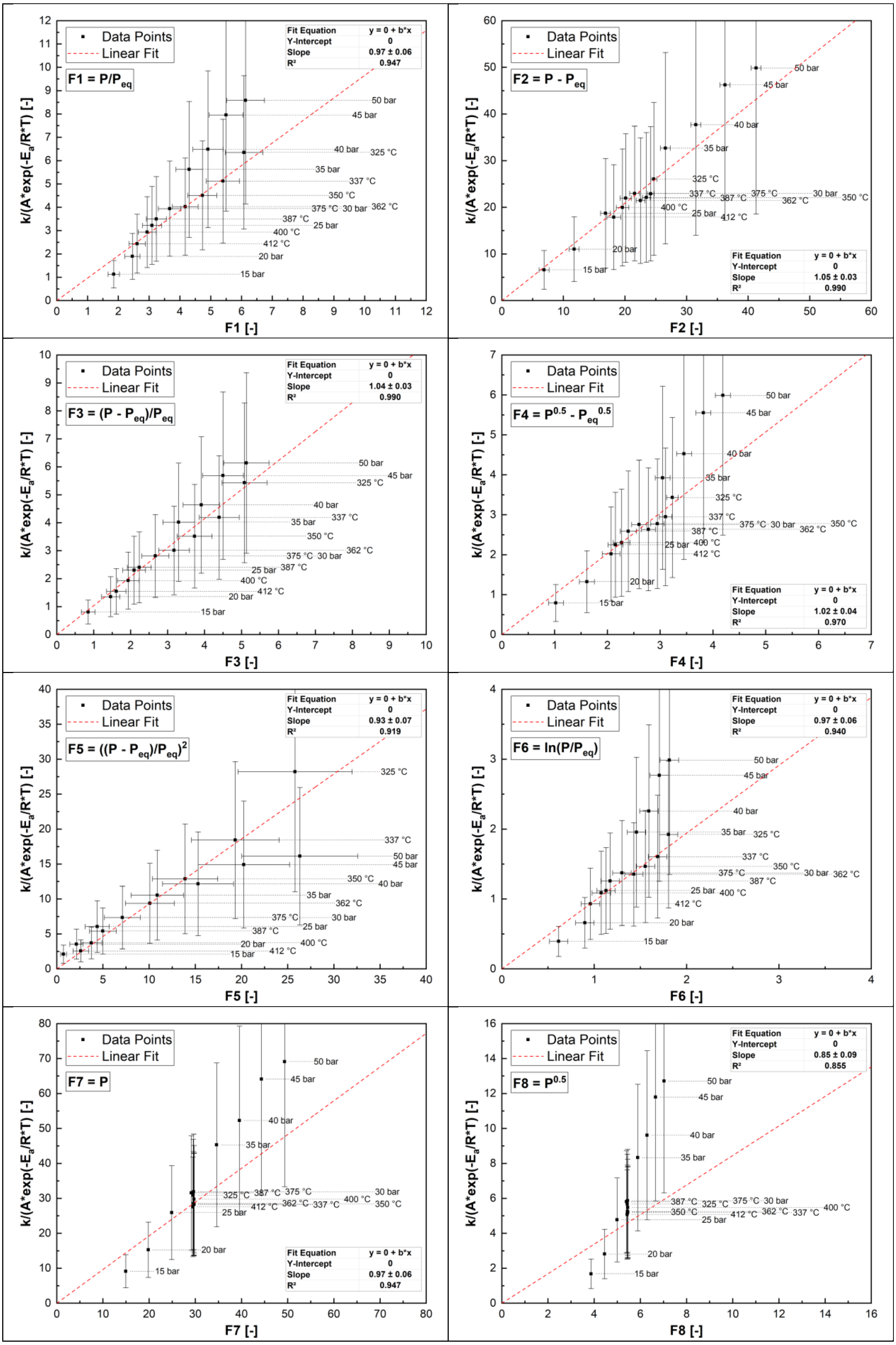

Data Points
Linear Fit
F1 = P/P$_{eq}$
Fit Equation y = 0 + b*x
Y-Intercept 0
Slope 0.97 ± 0.06
R² 0.947
F1 [-]
F2 = P - P$_{eq}$
Slope 1.05 ± 0.03
R² 0.990
F2 [-]
F3 = (P - P$_{eq}$)/P$_{eq}$
Slope 1.04 ± 0.03
R² 0.990
F3 [-]
F4 = P$^{0.5}$ - P$_{eq}$$^{0.5}$
Slope 1.02 ± 0.04
R² 0.970
F4 [-]
F5 = ((P - P$_{eq}$)/P$_{eq}$)$^2$
Slope 0.93 ± 0.07
R² 0.919
F5 [-]
F6 = ln(P/P$_{eq}$)
Slope 0.97 ± 0.06
R² 0.940
F6 [-]
F7 = P
Slope 0.97 ± 0.06
R² 0.947
F7 [-]
F8 = P$^{0.5}$
Slope 0.85 ± 0.09
R² 0.855
F8 [-]
k/(A*exp(-E$_a$/R*T) [-]

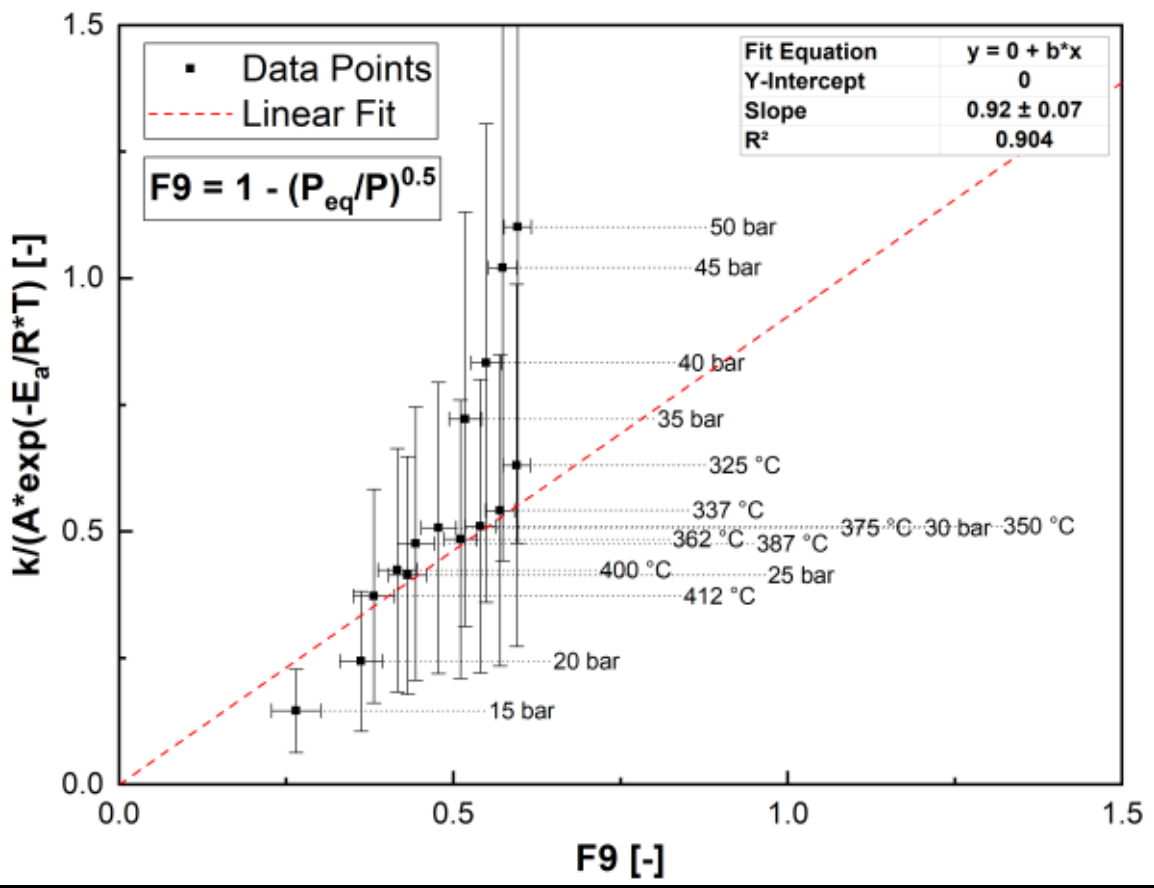

**Figure S6 – Linear fit of the calculated values for Li-RHC with 0.05 $TiCl_3$ for each of the driving force expressions listed in Table 3. Kinetic experiments were performed in two series, i.e., with different pressures (at 375 °C) and under same $H_2$-pressure (30 bar) with different temperatures. For more information, see Figure 3.b) of the main manuscript text.**

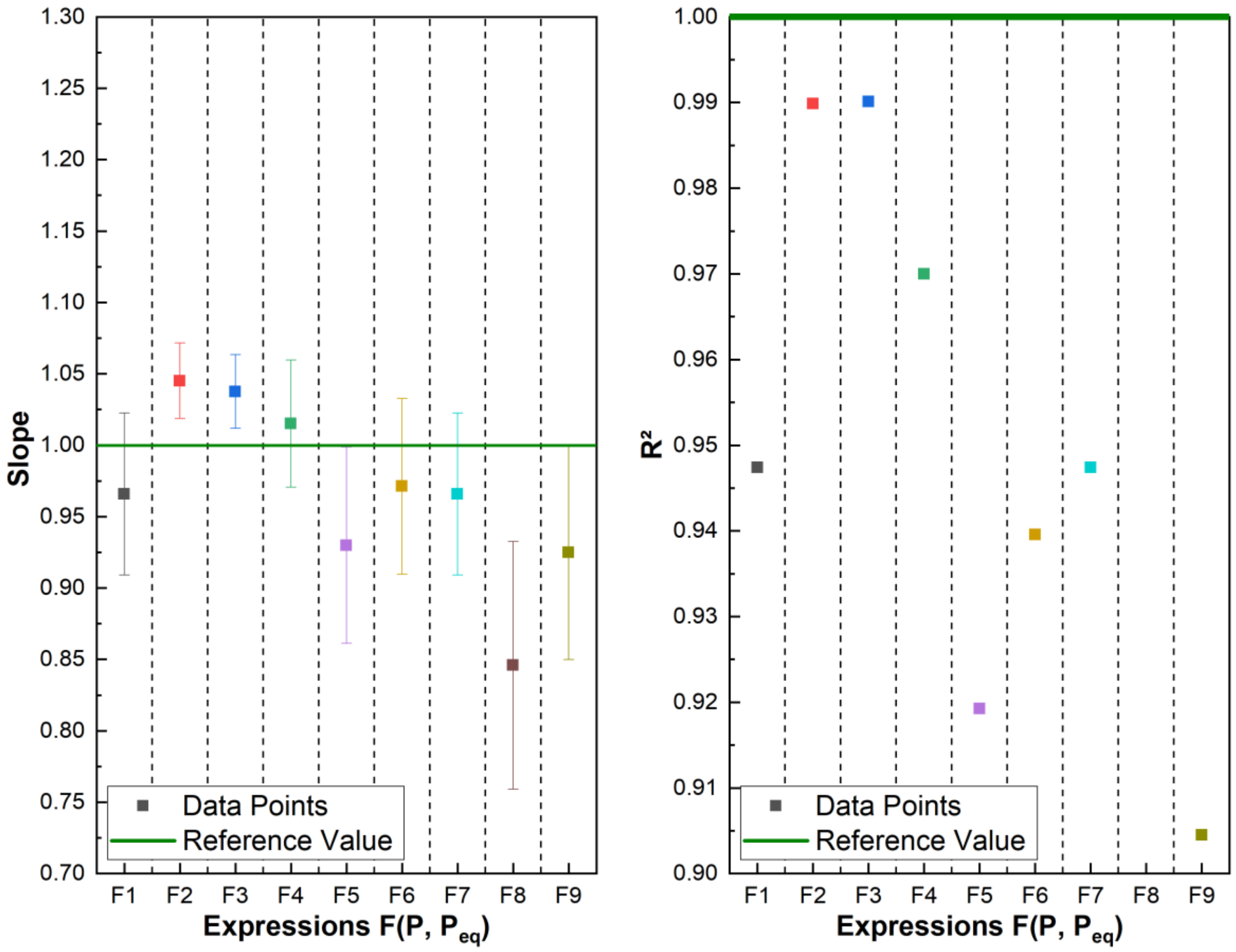

**Figure S7 – Results for the fit parameters for every tested driving force expression in regards to slope and coefficient of determination $R^2$.**

## Isokinetic Lines

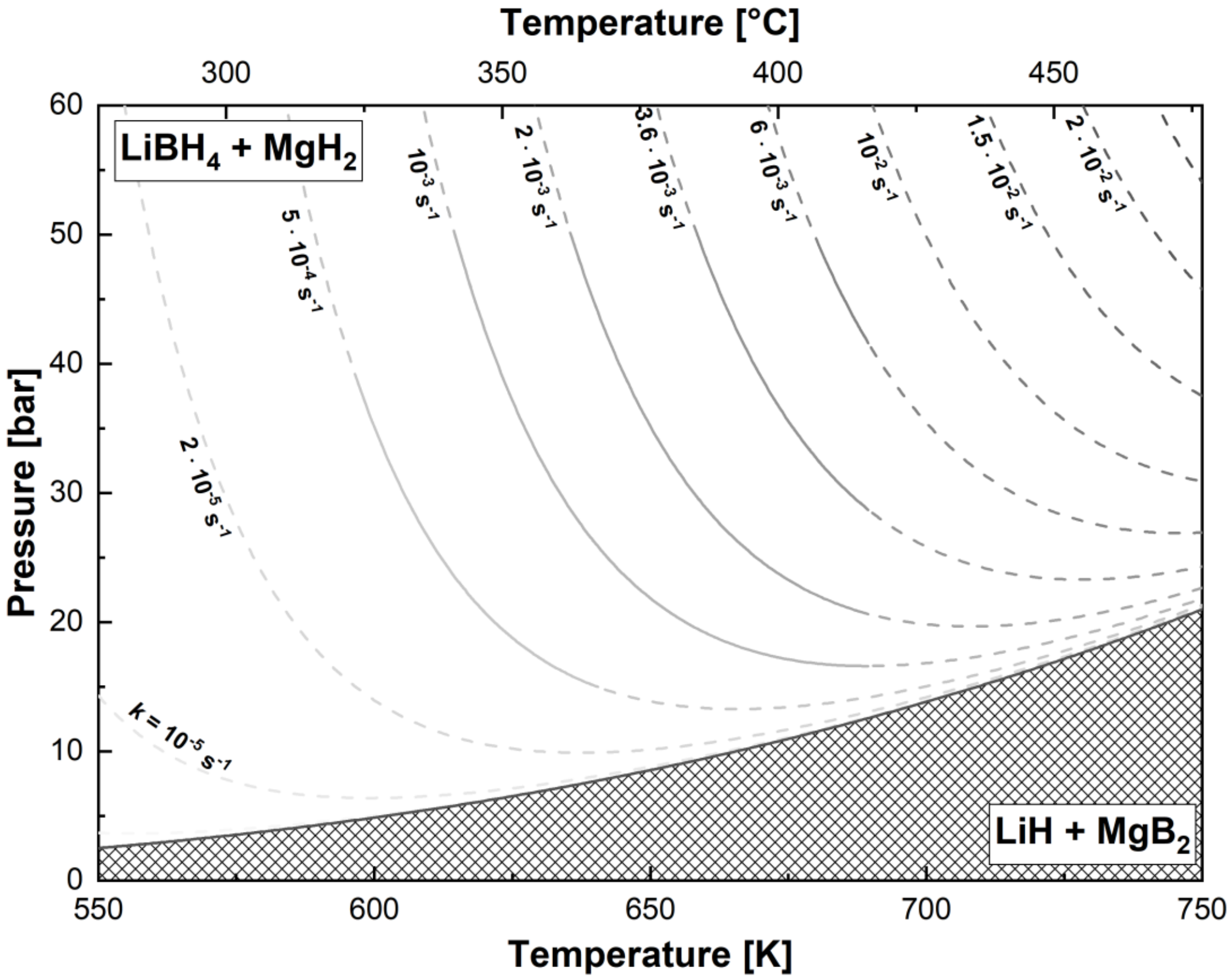

**Figure S8 – Thermodynamic stability regions and isokinetic contour lines for the Li-RHC system under absorption conditions. The isokinetic contour lines represent the conditions of pressure and temperature for which the calculated kinetic constant k is the same. The solid contour lines are in the region in which the model is considered to represent reliably the kinetic behavior of the system, while the dashed lines are result of the extrapolation beyond this region. The values shown near each of the lines are given for each k in $s^{-1}$.**